\documentclass[twocolumn, nofootinbib, aps, prd, 10pt]{revtex4-1}
\usepackage{graphicx, amsmath, amssymb, hyperref, bm, url}

\begin{document}

\title{Chasing the phantom: A closer look at Type Ia supernovae and the dark energy equation of state}

\author{Daniel L. Shafer}
\email{dlshafer@umich.edu}

\author{Dragan Huterer}
\email{huterer@umich.edu}

\affiliation{Department of Physics, University of Michigan, 450 Church St, Ann Arbor, MI 48109-1040}

\begin{abstract}

Some recent observations provide $> 2\sigma$ evidence for phantom dark energy -- a value of the dark energy equation of state less than the cosmological-constant value of $-1$. We focus on constraining the equation of state by combining current data from the most mature geometrical probes of dark energy: Type Ia supernovae (SNe Ia) from the Supernova Legacy Survey (SNLS3), the Supernova Cosmology Project (Union2.1), and the Pan-STARRS1 survey (PS1); cosmic microwave background measurements from Planck and WMAP9; and a combination of measurements of baryon acoustic oscillations. The combined data are consistent with $w = -1$ for the Union2.1 sample, though they present moderate ($\sim 1.9\sigma$) evidence for a phantom value when either the SNLS3 or PS1 sample is used instead. We study the dependence of the constraints on the redshift, stretch, color, and host galaxy stellar mass of SNe, but we find no unusual trends. In contrast, the constraints strongly depend on any external $H_0$ prior: a higher adopted value for the direct measurement of the Hubble constant ($H_0 \gtrsim 71~\text{km/s/Mpc}$) leads to $\gtrsim 2\sigma$ evidence for phantom dark energy. Given Planck data, we can therefore make the following statement at $2\sigma$ confidence: either the SNLS3 and PS1 data have systematics that remain unaccounted for or the Hubble constant is below 71~km/s/Mpc; else the dark energy equation of state is indeed phantom.

\end{abstract}

\maketitle

\section{Introduction} \label{sec:intro}

A key question in understanding the mechanism behind the acceleration of the Universe is the value of the dark energy equation of state, the ratio of pressure to energy density for dark energy: $w \equiv p_{_\text{DE}}/\rho_{_\text{DE}}$. Measurements so far \cite{Davis:2007na, Riess:2006fw, Knop:2003iy, Astier:2005qq, Riess:2006fw, WoodVasey:2007jb, Davis:2007na, Rubin:2008wq, Freedman:2009vv, Kowalski:2008ez, Hicken:2009dk, Kessler:2009ys, Amanullah:2010vv, Suzuki:2011hu} have generally been in good, even excellent, agreement with $w = -1$, the value corresponding to the vacuum energy density described by the famous cosmological-constant term in Einstein's equations of general relativity. Any measured departure from this value would not only profoundly shake up our understanding of the Universe, but also provide an important hint in our quest to understand cosmic acceleration.

It is therefore particularly important to measure the equation of state and search for any evidence of its variation in time. Over the past decade, there were several clear instances in which the measurements indicated that $w < -1$ at $\gtrsim 2\sigma$ evidence \cite{Huterer:2004ch, Alam:2006kj, Nesseris:2006ey, Zhao:2009ti}, though eventually these departures either were explained by known systematics in the data or quietly went away as new and better data became available. More recently, with the release of the first results from Planck \cite{Ade:2013zuv}, other such claims have been presented, such as \cite{Rest:2013bya}, which features high-quality data and a careful analysis including systematic errors \cite{Scolnic:2013aya} (see also \cite{Cheng:2013csa, Xia:2013dea}). This motivates us to investigate the current data in some detail, concentrating especially on the value of $w$ marginalized over other cosmological parameters.

The principal tool for studying the equation of state is the combination of three of the most mature probes of dark energy: Type Ia supernovae (SNe Ia), baryon acoustic oscillations (BAO), and cosmic microwave background (CMB) anisotropies. SNe Ia and BAO probe expansion at low and intermediate redshifts and are thus a crucial ingredient in studying dark energy. The CMB measurements effectively probe a single high-redshift distance (specifically, the distance to the surface of last scattering), which is crucial mainly because it provides complementary information to break degeneracy in the $\Omega_m$--$w$ plane. For comprehensive reviews of dark energy probes, see \cite{Frieman:2008sn, Weinberg:2012es}.

The rest of the paper is organized as follows. In Sec.~\ref{sec:data}, we describe the SN Ia, BAO, and CMB data that we use in our analysis. In Sec.~\ref{sec:results}, we present our results for the constraints on a constant dark energy equation of state along with several further analyses that were performed. In Sec.~\ref{sec:conclude}, we summarize and discuss our findings.

\section{Data Sets} \label{sec:data}

We begin by describing the data sets used in this analysis. We have used the three most mature probes of dark energy: SNe Ia, BAO, and CMB anisotropies. We focus on these three probes since they remain the most mature, well-studied, and robust dark energy probes at present. Furthermore, they are expected to be statistically independent for all practical purposes. Finally, being purely geometric in nature, they measure dark energy only through its effect on expansion history; therefore they are understood intuitively and may bypass certain systematic effects, such as those involved in growth of structure measurements, which are not very well understood.

\subsection{SN Ia data} \label{sec:data_SN}

SNe Ia were used to discover dark energy \cite{Riess:1998cb, Perlmutter:1998np} and still provide the best constraints on dark energy. SNe Ia are very bright standard candles that are useful for measuring cosmic distances.

SNe Ia constrain cosmology by providing essentially one measurement each of the luminosity distance $D_L(z) = (1 + z) \, r(z)$ to the redshift of the SN. The theoretical apparent magnitude is then given by
\begin{equation}
m_\text{th}(z) = 5 \log_{10} \left(\frac{H_0}{c} \, D_L(z) \right) + \mathcal{M} \ ,
\label{eq:mth}
\end{equation}
where the constant magnitude offset $\mathcal{M}$ is a nuisance parameter that depends on both the absolute magnitude of a SN Ia and the Hubble constant $H_0$. It has long been known that there exist useful correlations between the peak absolute magnitude of a SN Ia and both the stretch (or broadness) and photometric color of its light curve. Simply put, a broader or bluer SN light curve corresponds to a brighter SN. Thus we compare the theoretical apparent magnitude with the measured magnitude after light-curve correction:
\begin{equation}
m_\text{corr} = m_B + \alpha \times (\text{stretch}) - \beta \times (\text{color}) \ ,
\label{eq:mcorr}
\end{equation}
where the stretch and color measures are specific to the light-curve fitter employed (e.g.\ SALT2 \cite{Guy:2007dv} or SiFTO \cite{Conley:2008xx}) and where $\alpha$ and $\beta$ are two additional nuisance parameters.

Recent work has concentrated on estimating correlations between measurements of individual SN Ia magnitudes as a way of accounting for the numerous systematic effects which must be controlled in order to improve SN Ia constraints significantly beyond their current level \cite{Ruiz:2012rc}. A complete covariance matrix for SNe Ia includes all identified sources of systematic error in addition to the intrinsic scatter and other sources of statistical error. The $\chi^2$ statistic is then given by
\begin{equation}
\chi^2 = \Delta \mathbf{m}^\intercal \mathbf{C}^{-1} \Delta \mathbf{m} \ ,
\label{eq:covchi2}
\end{equation}
where $\Delta \mathbf{m} = \mathbf{m}_\text{corr} - \mathbf{m}_\text{th}(\mathbf{p})$ is the vector of differences between the observed, corrected magnitudes $\mathbf{m}_\text{corr}$ of $N$ SNe Ia and the theoretical predictions $\mathbf{m}_\text{th}(\mathbf{p})$ that depend on the set of cosmological parameters $\mathbf{p}$. Here, $\mathbf{C}$ is the $N \times N$ covariance matrix between individual SNe.

In this analysis, we compare current SN Ia data from three separate compilations: the Union2.1 compilation from the Supernova Cosmology Project, the three-year compilation from the Supernova Legacy Survey (SNLS3), and the compilation of the first SN sample from the Pan-STARRS1 survey (PS1).

\subsubsection{Union2.1} \label{sec:Union2.1}

The Union2.1 compilation \cite{Suzuki:2011hu} from the Supernova Cosmology Project (\url{http://supernova.lbl.gov/Union/}) improves on the previous Union2 compilation \cite{Amanullah:2010vv} by introducing 27 additional SNe at high redshift, making it both the largest compilation (580~SNe) and the one with the most high-redshift SNe ($\sim$30 at $z \gtrsim 1$). The compilation combines several different samples in each redshift region (low, intermediate, and high), making the redshift coverage very complete but also making the compilation very inhomogeneous.

For this analysis, we include all identified systematic errors via the covariance matrix provided. The SN magnitudes have been pre-corrected for stretch and color using best-fit values for $\alpha$ and $\beta$, and we have verified that our SN-only constraints match those presented in \cite{Suzuki:2011hu}.

\subsubsection{SNLS3} \label{sec:SNLS3}

Results from the first three years of the Supernova Legacy Survey include measurements of $\sim$250 SNe at intermediate-to-high redshifts. When combined with $\sim$100 low-redshift SNe, $\sim$100 SNe from the Sloan Digital Sky Survey (SDSS), and $\sim$10 high-redshift SNe from the Hubble Space Telescope, they produce a compilation \cite{Conley:2011ku} of 472~SNe with good redshift coverage out to $z \sim 1$ and some SNe extending to $z \simeq 1.4$. The SNLS3 compilation contains the largest homogeneous sample and includes many of the best-measured SNe along with a detailed analysis of systematic errors \cite{Conley:2011ku}. Note that the SNLS3 and SDSS samples have been recalibrated \cite{Betoule:2012an} and that new cosmological results, including the full SDSS sample, are forthcoming.

We use the SN Ia data and covariance matrices provided (\url{https://tspace.library.utoronto.ca/snls}) to compute the full covariance matrix, which includes all identified sources of statistical and systematic error. Like the corrected SN magnitudes, the covariance matrix is a function of the light-curve nuisance parameters $\alpha$ and $\beta$. For practical reasons, and for a fairer comparison with other SN Ia data sets, we fix these parameters at their best-fit values ($\alpha = 1.43$, $\beta = 3.26$) throughout the analysis. It is worth noting that completely marginalizing over these parameters (varying them both when computing the corrected magnitudes and when building the covariance matrix) has a negligible effect on constraints in our parameter space. We verify that our SN-only constraints match those in \cite{Conley:2011ku, Ruiz:2012rc}, where $\alpha$ and $\beta$ are varied.

It is important to note that, in constraints with SNLS3, we follow the prescription in \cite{Conley:2011ku} and marginalize over a model with two distinct $\mathcal{M}$ parameters, where a mass cut ($10^{10} M_\odot$) of the host galaxy dictates which $\mathcal{M}$ applies. This is meant to correct for environmental dependencies of SN Ia magnitudes on host galaxy properties and is empirical in nature. We discuss and investigate this issue further in Sec.~\ref{sec:hostmass}.

\subsubsection{PS1} \label{sec:PS1}

The primary goal of the Panoramic Survey Telescope and Rapid Response System (Pan-STARRS) is to detect Solar System objects by making precise, repeated observations of a wide field of view. These observations also lead to the discovery of SNe Ia, which can be spectroscopically confirmed in follow-up observations. Recently-published SN results from the first 1.5 years of the Pan-STARRS1 Medium Deep Survey include a compilation \cite{Rest:2013bya} of 313~SNe, 112 of which were discovered via Pan-STARRS. The rest (201~SNe) come from a combination of low-redshift samples. Aside from the low-redshift anchor, all SNe come from the same instrument, making this compilation very homogeneous. A full systematic analysis is described in \cite{Scolnic:2013aya}. Due to the smaller number of SNe and the lack of high-redshift SNe, the PS1 compilation is not competitive with current constraints from Union2.1 or SNLS3, but the survey is ongoing and this will eventually change. The current compilation nevertheless provides good constraints on a constant-$w$ model of dark energy when combined with other probes, so we study it here.

We use the SN Ia data provided (\url{http://wachowski.pha.jhu.edu/~dscolnic/PS1_public/}) and adopt the covariance matrix to account for all identified systematic errors. As with Union2.1, the SN magnitudes have been pre-corrected for stretch and color, and we have verified that our SN-only constraints agree with those in \cite{Rest:2013bya}.

\subsection{BAO data} \label{sec:data_BAO}

Baryon acoustic oscillations (BAO) are the regular, periodic fluctuations of visible matter density in large-scale structure (LSS) resulting from sound waves propagating in the early Universe. In recent years, measurements of BAO have proven to be useful geometric probes of dark energy. A measurement of the position of the BAO feature in the LSS power spectrum or correlation function basically provides a precise measurement of a spherically-averaged comoving distance to the effective redshift of the survey. New measurements over a wide range of redshifts are making it possible to map expansion history with the BAO distance, analogous to the way SNe Ia map expansion with luminosity distance. For our BAO constraints, we combine recent measurements of the BAO feature from the Six-degree-Field Galaxy Survey (6dFGS) \cite{Beutler:2011hx}, the SDSS Luminous Red Galaxies (SDSS LRG) \cite{Padmanabhan:2012hf}, and the SDSS-III DR9 Baryon Oscillation Spectroscopic Survey (BOSS) \cite{Anderson:2012sa}.

Different authors report their measurement of the BAO feature using different distilled observable quantities. The surveys included here report constraints on the quantity $r_s(z_d)/D_V(z)$ or its inverse, where $r_s(z_d)$ is the comoving sound horizon at the redshift of the baryon drag epoch and $D_V$ is a spherically-averaged (two tangential and one radial) distance measure \cite{Eisenstein:2005su} given by
\begin{equation}
D_V(z) \equiv \left[(1 + z)^2 \, D_A^2(z) \frac{cz}{H(z)} \right]^{1/3} ,
\label{eq:DV}
\end{equation}
where $D_A(z) = r(z)/(1 + z)$ is the angular diameter distance. We compute the sound horizon via
\begin{align}
r_s(z) &= \int_0^t \frac{c_s}{a} \, \text{d}t' = \int_0^a \frac{c_s}{a'{}^2 H(a')} \, \text{d}a' \ , \\
c_s &= \frac{c}{\sqrt{3 \, (1 + R)}} \ , \nonumber
\end{align}
where the sound speed $c_s$ depends on the ratio of baryon energy density to photon energy density, which is proportional to the scale factor:
\begin{equation}
R \equiv \frac{3 \rho_b}{4 \rho_\gamma} \approx 31500 \, \Omega_b h^2 \, \left(\frac{T_\text{CMB}}{2.7~\text{K}} \right)^{-4} a \ .
\end{equation}
The redshift of the baryon drag epoch is given by the fitting formula \cite{Eisenstein:1997ik}
\begin{equation}
z_d = \frac{1291 \, (\Omega_m h^2)^{0.251}}{1 + 0.659 \, (\Omega_m h^2)^{0.828}} \left[1 + b_1 (\Omega_b h^2)^{b_2} \right] \ ,
\end{equation}
where
\begin{align*}
b_1 &= 0.313 \, (\Omega_m h^2)^{-0.419} \left[1 + 0.607 \, (\Omega_m h^2)^{0.674} \right] \ , \\
b_2 &= 0.238 \, (\Omega_m h^2)^{0.223} \ .
\end{align*}
It is important to include a term for radiation in $H(a)$. One can write $\Omega_r = \Omega_m a_\text{eq}$, where $a_\text{eq} = 1/(1 + z_\text{eq})$ is the scale factor at the epoch of matter-radiation equality and $z_\text{eq}$ is approximated by
\begin{equation}
z_\text{eq} \approx 25000 \, \Omega_m h^2 \, \left(\frac{T_\text{CMB}}{2.7~\text{K}} \right)^{-4}.
\end{equation}
We assume the value $T_\text{CMB} = 2.7255~\text{K}$ in our analysis.

The measured values of the BAO parameters are summarized in Table~\ref{tab:baovals}. Covariance between different surveys should be negligible here, so we treat these as independent measurements.

Note that previous measurements of the BAO feature from SDSS LRG data (e.g.\ \cite{Eisenstein:2005su, Percival:2009xn}) cannot be used simultaneously with the measurement from \cite{Padmanabhan:2012hf} since roughly the same galaxy sample is analyzed. The measurement from \cite{Padmanabhan:2012hf} makes use of a reconstruction technique to enhance the BAO signal and increase the precision of the distance measurement. We use this measurement since it is the most precise and avoids the correlation between the pair of measurements from \cite{Percival:2009xn}, where the SDSS LRG sample is combined with the SDSS main galaxy sample.

Also note that we choose to leave out the BAO measurements from the WiggleZ Dark Energy Survey \cite{Blake:2011en}, which measures the BAO distance in three redshift slices ($z_\text{eff} = 0.44, 0.6, 0.73$). These measurements are somewhat correlated with the BOSS measurement due to overlap in sky area and redshift, and so far no correlation coefficients have been estimated. Although the correlation is probably negligible (in part due to shot noise), adding the WiggleZ measurements to the other BAO measurements improves our constraints only very slightly, so we leave them out.

\begin{table}[h]
\begin{center}
\setlength{\tabcolsep}{0.5em}
\begin{tabular}{|| c | c | c | c ||}
\hline \hline
\rule[-3mm]{0mm}{8mm} Sample 	 & $z_\text{eff}$ & Parameter									   & Measurement       \\ \hline \hline
\rule[-3mm]{0mm}{8mm} 6dFGS 	 & $0.106$ 			  & $r_s(z_d)/D_V(z_\text{eff})$ & $0.336 \pm 0.015$ \\ \hline
\rule[-3mm]{0mm}{8mm} SDSS LRG & $0.35$ 				& $D_V(z_\text{eff})/r_s(z_d)$ & $8.88  \pm 0.17$  \\ \hline
\rule[-3mm]{0mm}{8mm} BOSS 		 & $0.57$         & $D_V(z_\text{eff})/r_s(z_d)$ & $13.67 \pm 0.22$  \\ \hline
\end{tabular}
\end{center}
\caption{Summary of BAO measurements combined in the analysis. We list the survey from which the measurement comes, the effective redshift of the survey, the observable parameter constrained, and its measured value.}
\label{tab:baovals}
\end{table}

\begin{figure*}[t]
\includegraphics[width=\textwidth]{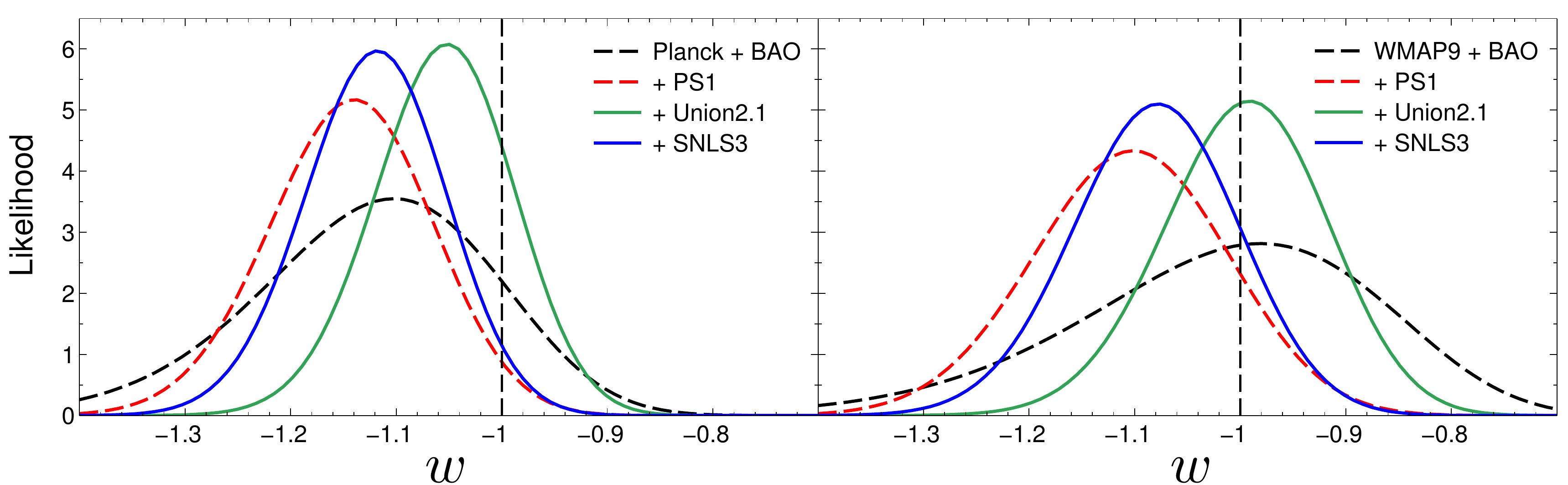}
\caption{Likelihood curves for a constant equation of state $w$ in a flat universe, using Planck CMB data (left panel) and WMAP9 CMB data (right panel). We compare constraints from CMB + BAO data alone (dashed black) to those which additionally include SN Ia data from SNLS3 (blue), Union2.1 (green), or PS1 (dashed red). All likelihoods are marginalized over other cosmological and nuisance parameters, as explained in the text.}
\label{fig:wlike}
\end{figure*}

\subsection{CMB data} \label{sec:data_CMB}

Although the CMB contains relatively little geometric information about dark energy, the position of the first peak in the power spectrum basically provides one very precise measurement of the angular diameter distance to recombination at $z \approx 1100$. This measurement helps break degeneracy between the dark energy equation of state and $\Omega_m$ \cite{Frieman:2002wi}. In our analysis, we include CMB constraints from Planck \cite{Ade:2013zuv} and also from WMAP9 \cite{Hinshaw:2012aka} for comparison.

We summarize CMB information using the following CMB observables:
\begin{align}
l_a &\equiv \pi \, (1 + z_\ast) \, \frac{D_A(z_\ast)}{r_s(z_\ast)} \label{eq:la} \ , \\
R &\equiv \frac{\sqrt{\Omega_m H_0^2}}{c} \, (1 + z_\ast) \, D_A(z_\ast) \label{eq:R} \ .
\end{align}
The redshift $z_\ast$ of decoupling is given by the fitting formula \cite{Hu:1995en}
\begin{align}
z_\ast = 1048 \, &\left[1 + 0.00124 \, (\Omega_b h^2)^{-0.738} \right] \\
&\times \left[1 + g_1 (\Omega_m h^2)^{g_2} \right] \ , \nonumber
\end{align}
where
\begin{align*}
g_1 &= \frac{0.0783 \, (\Omega_b h^2)^{-0.238}}{1 + 39.5 \, (\Omega_b h^2)^{0.763}} \ , \\
g_2 &= \frac{0.560}{1 + 21.1 \, (\Omega_b h^2)^{1.81}} \ .
\end{align*}

Following \citet{Wang:2013mha}, we use the Markov chains from the Planck Legacy Archive (PLA) to derive constraints on the parameter combination $(l_a \, , R \, , z_*)$, which is known to efficiently summarize CMB information on dark energy, with the measurements themselves independent of the dark energy model to a good approximation. We assume the same model that we constrain in this analysis (flat universe, constant $w$) when deriving the CMB observables. For the Planck data, we use information from the temperature power spectrum combined with WMAP polarization at low multipoles (Planck + WP). We also use the PLA chains to derive the corresponding measurements for WMAP9 (temperature and polarization data) with the same model assumptions. The CMB measurements are summarized in Table~\ref{tab:cmbvals}.

\begin{table}[b]
\begin{center}
\setlength{\tabcolsep}{1em}
\begin{tabular}{|| c || c | c | c ||}
\hline \hline
\rule[-3mm]{0mm}{8mm} $\bar{x} \pm \, \sigma$ & Planck               & WMAP9                \\ \hline \hline
\rule[-3mm]{0mm}{8mm} $l_a$ 	                & $301.65  \pm 0.18$   & $301.98  \pm 0.66$   \\ \hline
\rule[-3mm]{0mm}{8mm} $R$ 	                  & $1.7499  \pm 0.0088$ & $1.7302  \pm 0.0169$ \\ \hline
\rule[-3mm]{0mm}{8mm} $z_*$                   & $1090.41 \pm 0.53$   & $1089.09 \pm 0.89$   \\ \hline
\end{tabular}
\end{center}
\caption{Mean values and standard deviations of the CMB measurements used in our analysis. The measurements for both Planck and WMAP9 were obtained using the Markov chains provided by the Planck collaboration. We assumed the model with a flat universe and constant dark energy equation of state, the same model we constrain in this analysis.}
\label{tab:cmbvals}
\end{table}

We evaluate the correlation matrix for $(l_a \, , R \, , z_*)$ for Planck to be
\[
\left(
\begin{array}{ccc}
1.0000  & 0.5262 & 0.4708 \\
0.5262  & 1.0000 & 0.8704 \\
0.4708  & 0.8704 & 1.0000 \\
\end{array}
\right).
\]
The same correlation matrix for WMAP9 is
\[
\left(
\begin{array}{ccc}
1.0000 & 0.4077 & 0.5132 \\
0.4077 & 1.0000 & 0.8580 \\
0.5132 & 0.8580 & 1.0000 \\
\end{array}
\right).
\]

We have explicitly verified that, when multiple probes are combined, the constraints on $w$ obtained directly from the CMB chains are in good agreement with results obtained using the measurements in Table~\ref{tab:cmbvals}. For the base case of Planck combined with BAO and SNLS3 SNe, the best-fit values of $w$ differ by less than 0.1$\sigma$. The discrepancy is greater when the complementary SN data are not included, but the difference is still less than 0.3$\sigma$ for the data combinations we consider.

Note that our measurements cannot be directly compared to those presented in \cite{Wang:2013mha} because of different assumptions: we do not include Planck lensing information, we assume a flat model with $w$ as a free parameter instead of a $\Lambda$ model with curvature, and we treat $z_*$ as an observable in place of $\Omega_b h^2$.

\begin{figure}[b]
\includegraphics[width=0.48\textwidth]{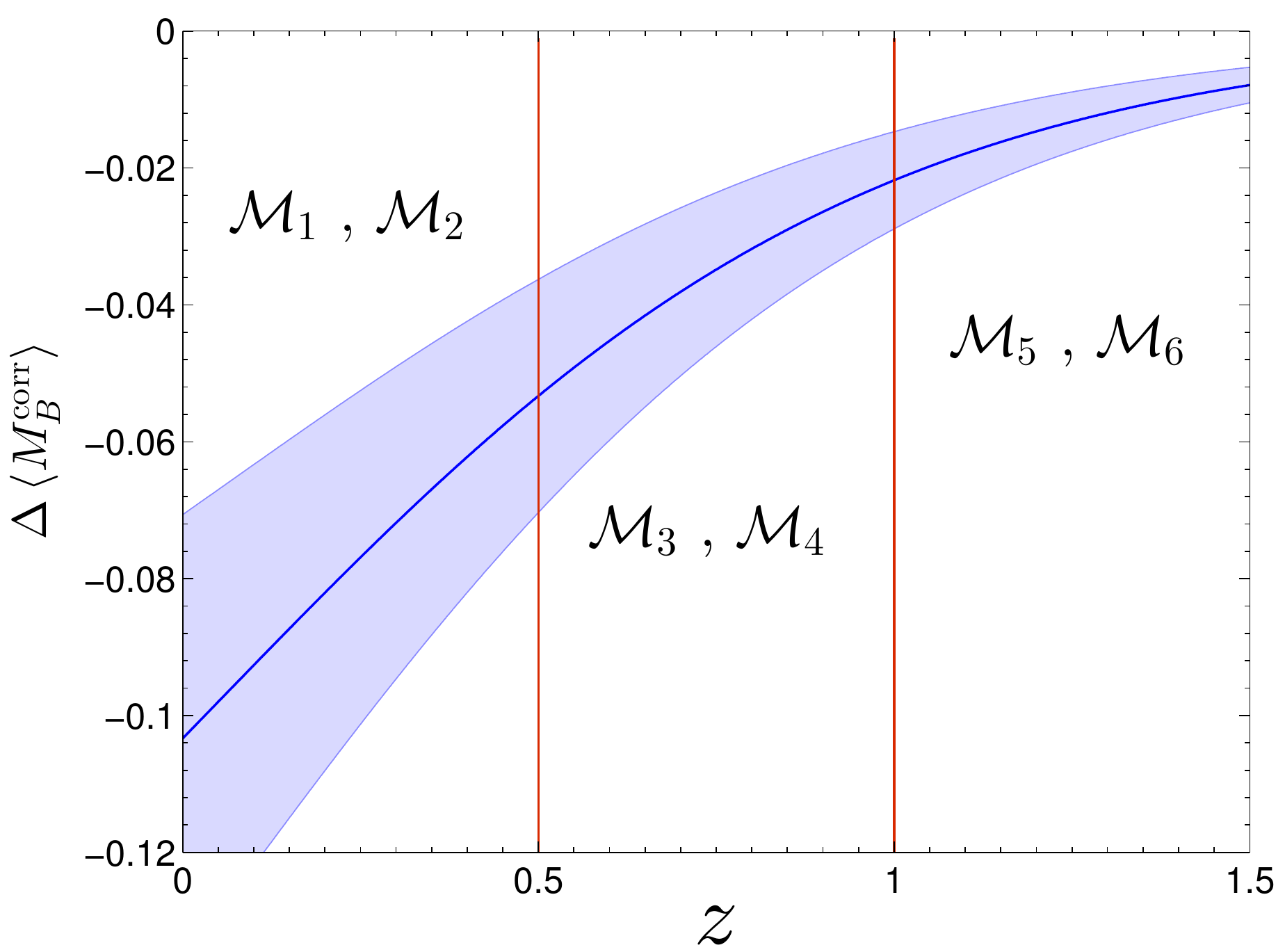}
\caption{Evolution of the mass step predicted from a toy model calibrated using data from the Nearby Supernova Factory. This is similar to Fig.~11 of \cite{Rigault:2013gux}, though we include a region of uncertainty by propagating errors in the mass step and local star-forming fraction measured at $z = 0.05$ from the Nearby Supernova Factory data. Vertical lines separate the three redshift bins, each of which contains two $\mathcal{M}$ nuisance parameters, one for each host galaxy mass range.}
\label{fig:mass_step}
\end{figure}

\begin{figure*}[t]
\includegraphics[width=0.48\textwidth]{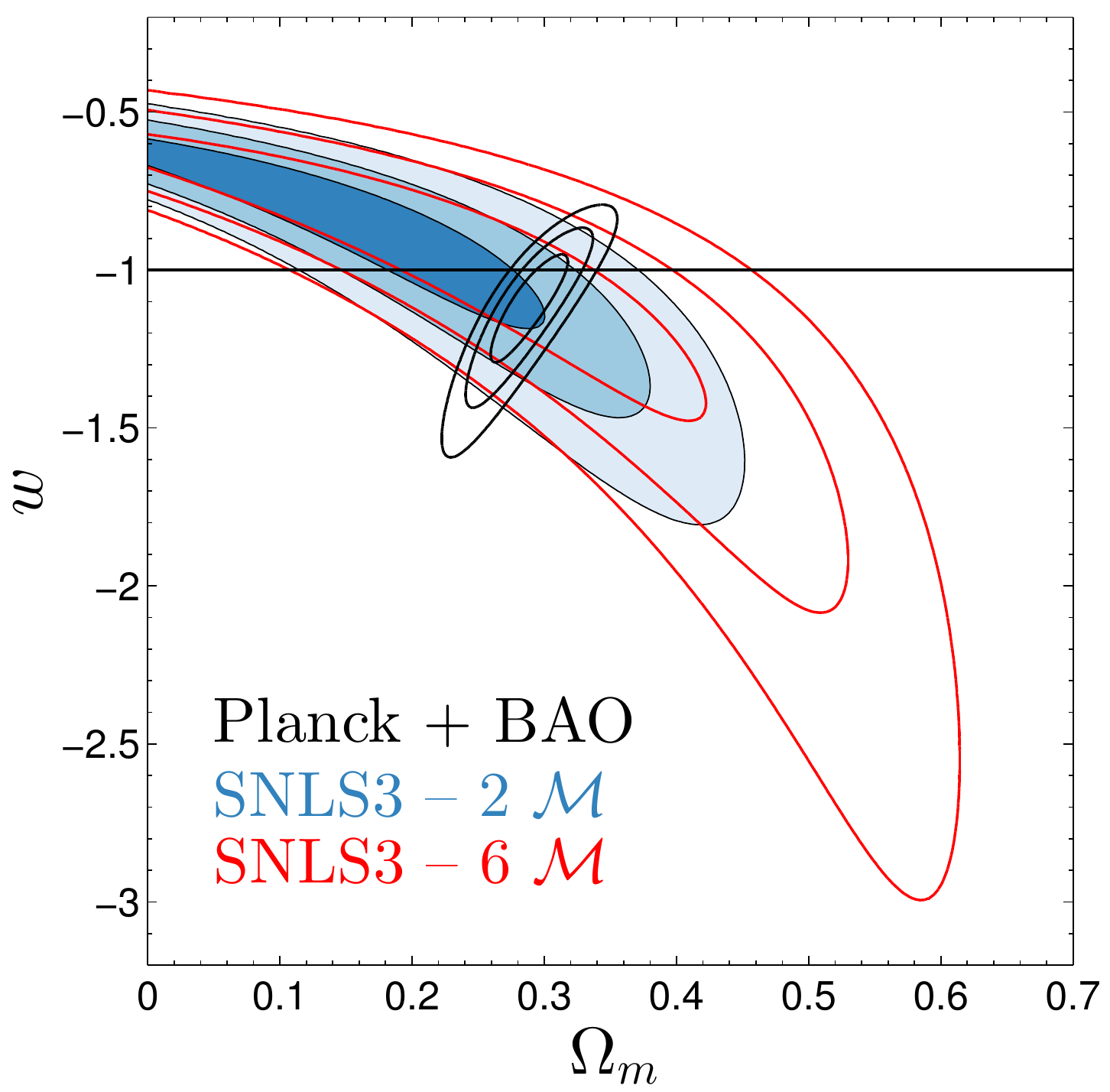}
\includegraphics[width=0.5\textwidth]{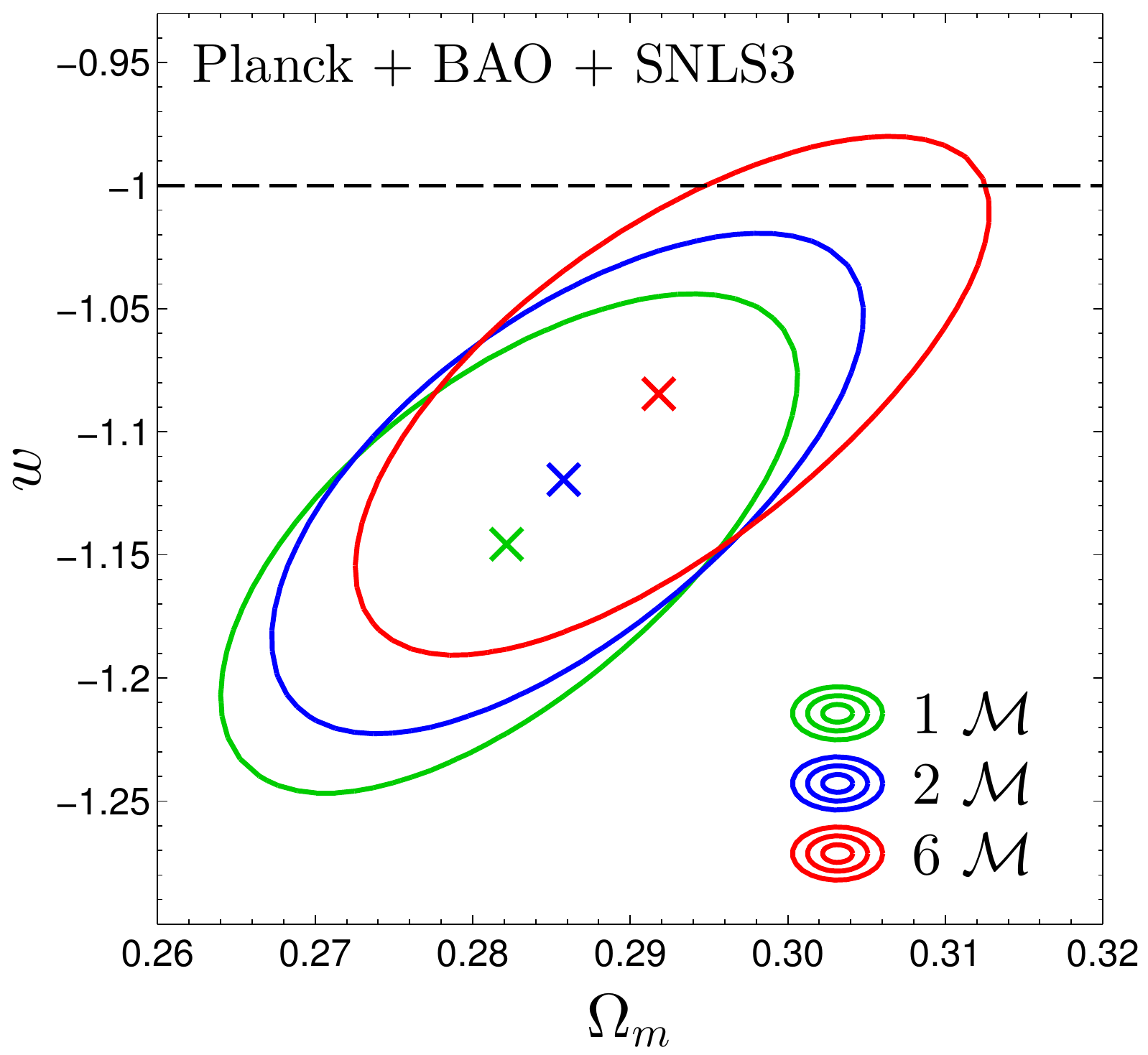}
\caption{Effect of allowing for evolution of the mass step in redshift bins in the SN Ia analysis. \textit{Left}: 68.3\%, 95.4\% and 99.7\% likelihood contours in the $\Omega_m$--$w$ plane for SNLS3 data analyzed the standard way with two $\mathcal{M}$ nuisance parameters (filled blue) and a new way with six $\mathcal{M}$ parameters (open red), one for each of two mass bins and three redshift bins. Planck + BAO constraints (open black) are overlaid for comparison. \textit{Right}: 68.3\% contours in the same plane for combined Planck + BAO + SNLS3 data using one, two, or six $\mathcal{M}$ parameters.}
\label{fig:six_M_Om_w}
\end{figure*}

\section{Results} \label{sec:results}

\subsection{Constraint methodology} \label{sec:method}
The complete parameter set used in our analysis is
\[
p_i \in \{\Omega_m \, , w \, , \Omega_m h^2 \, , \Omega_b h^2 \, , \{\mathcal{M}_i\} \} \ ,
\]
where we marginalize over $\Omega_m h^2$ and $\Omega_b h^2$ for the CMB and BAO constraints and over one or more SN Ia nuisance parameters $\mathcal{M}_i$ (see Secs.~\ref{sec:SNLS3}~and~\ref{sec:hostmass}). Given the small number of parameters, we calculate constraints using brute-force computation of likelihoods over a grid of parameter values. We assume a Gaussian likelihood $\mathcal{L} \propto \exp(-\chi^2/2)$, where we have ignored the $1/\sqrt{\det \mathbf{C}}$ prefactor, which is a constant and thus cancels out in likelihood ratios. Note that, in general, the SN covariance matrix is a function of the SN nuisance parameters. If we were to vary those parameters, we would need to recompute the SN covariance matrix at each step; however, one might still want to drop the Gaussian prefactor, as it can bias the values of recovered parameters if included (e.g.\ \cite{Kelly:2007jy} and Appendix~B of \cite{Conley:2011ku}). Finally, note that aside from the implicit prior that $\{\Omega_m \, , \Omega_m h^2 \, , \Omega_b h^2 \} \geq 0$, we assume flat priors on all of the parameters.

\subsection{Basic constraints} \label{sec:constraints}

Combined constraints on the equation of state, marginalized over the other parameters, are shown in Fig.~\ref{fig:wlike}, where the left panel shows the Planck data combined with the BAO and SN Ia data, while the right panel shows the same for WMAP9. CMB and BAO data alone constrain the equation of state rather weakly. With Planck, there is a preference for $w < -1$, but at $\simeq 1\sigma$ it is not significant. There is no preference at all with WMAP9. Note also that the constraints with Planck are visibly better than those with WMAP9, as Planck measures all of the CMB distance parameters $(l_a \, , R \, , z_*)$ more precisely, with errors that are 2--3 times smaller.

Things get more interesting when SN Ia data are added. The Union2.1 data set produces the best constraints when combined with CMB and BAO, marginally better than the constraints with SNLS3. Again, though, this leads to good agreement with a cosmological constant, with an insignificant preference for $w < -1$ driven by the Planck data. However, when SNLS3 or PS1 data are used, we find a preference for $w < -1$ at the $1.8\sigma$ (SNLS3) or $1.9\sigma$ (PS1) level with Planck and the $1\sigma$ (SNLS3) or $1.2\sigma$ (PS1) level with WMAP9.\footnote{Since the posterior likelihoods are not perfectly Gaussian, we always determine $\sigma$ values by computing the integral of the likelihood between the two values of $w$ where the likelihood equals that at $w = -1$. The quoted multiple of $\sigma$ is the number of standard deviations that enclose this probability in a \textit{Gaussian} distribution.} Note that the PS1 data give slightly stronger evidence for $w < -1$, even though the overall constraints are weaker.

It is useful to study the SN Ia constraints in more detail, which we do in the following two subsections. Our work here complements the detailed systematic analyses in \cite{Sullivan:2010mg, Conley:2011ku, Gupta:2011pa, Ruiz:2012rc, Kessler:2012gn, Johansson:2012si, Childress:2013xna, Wang:2013yja, Rigault:2013gux, Scolnic:2013aya}. The particular focus of this paper is the effect of potential SN Ia systematics and external priors on evidence for ``phantom'' behavior of dark energy where $w < -1$.

\begin{figure*}[t]
\includegraphics[width=0.8\textwidth]{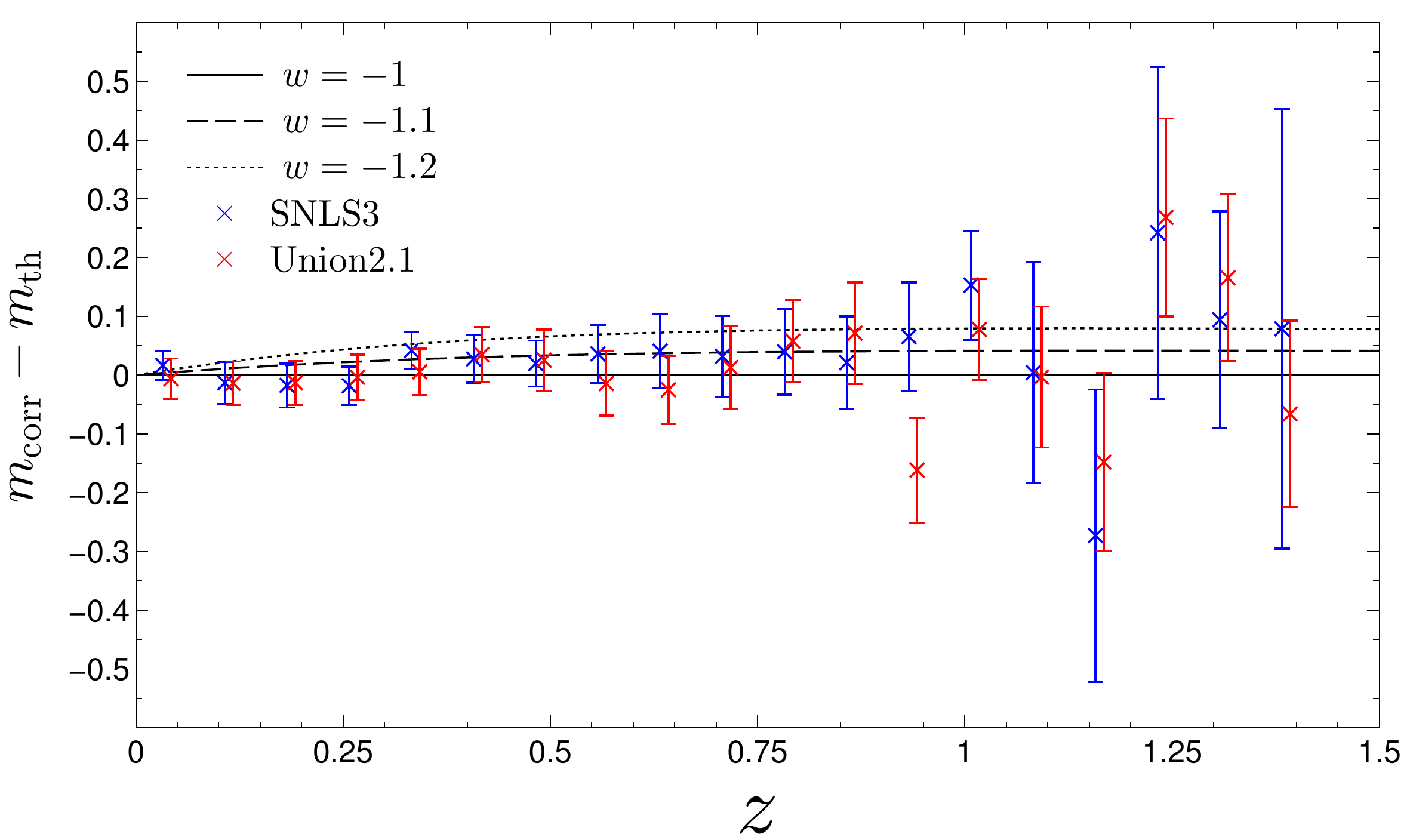}
\caption{Residuals of SN Ia magnitudes, binned by redshift (inverse-covariance weights), for SNLS3 (blue) and Union2.1 (red). All curves and data points are relative to a flat $\Lambda$CDM cosmology with $\Omega_m = 0.3$, which is roughly the best-fit value from CMB and BAO data. The plot shows the degree to which SNe in each redshift range pull toward $w < -1$, and we show several theory curves with constant $w$ for comparison.}
\label{fig:residuals}
\end{figure*}

\subsection{SN Ia host mass correction} \label{sec:hostmass}

Recently, much work has been focused on understanding the environmental dependence of SNe Ia, which presumably is not only one of the important factors contributing to intrinsic scatter of SN magnitudes but also an important systematic effect. In particular, there is evidence that the absolute magnitude of SNe Ia is correlated with host galaxy properties such as specific star formation rate, metallicity, and stellar mass \textit{after} the usual light-curve stretch and color correction \cite{Sullivan:2010mg, Gupta:2011pa, Johansson:2012si, Childress:2013xna, Rigault:2013gux, Scolnic:2013aya}. Most striking is the evidence for a ``mass step'' where SNe in more massive hosts ($\gtrsim 10^{10} M_\odot$) are brighter, on average, after light-curve correction. This is consistent with a step function, suggesting that one could fit for two separate magnitude offsets (i.e.\ $\mathcal{M}$), one for SNe in low-mass hosts and one for SNe in high-mass hosts. Indeed this was prescribed for SNLS3 in \cite{Conley:2011ku}.

Of course, the mass of the host galaxy itself should have no direct physical influence on SN luminosity, so something else must be at work. Recent measurements from Nearby Supernova Factory data \cite{Rigault:2013gux} have indicated strong ($\sim 3.1\sigma$) evidence that SNe Ia in \textit{locally} passive environments are brighter on average than those in active star-forming environments. The authors further show that this can explain the observed mass step, as passive environments are more common in high-mass galaxies. This is especially important because the fraction of SNe Ia in locally star-forming environments surely evolves with redshift, and therefore the amplitude of the mass step should also evolve. This is a systematic effect not corrected for by the introduction of two $\mathcal{M}$ parameters, and the authors estimate a bias on the equation of state of $\Delta w \simeq 0.06$.

Fig.~\ref{fig:mass_step} shows a toy model for the redshift evolution of the mass step from the analysis of \cite{Rigault:2013gux}, with errors that we have estimated by propagating errors in the mass step and local star-forming fraction measured at $z = 0.05$ from the Nearby Supernova Factory data. Given the astrophysical uncertainties in linking the star formation rate to host stellar mass and the latter to absolute magnitude of SNe Ia, we do not try to use any fixed model to correct for this. Instead, we use a less model-dependent parametrization of the relation between the observed host galaxy mass and absolute magnitude by allowing for two independent values of $\mathcal{M}$ in each of three redshift bins: $z \leq 0.5$, $0.5 < z \leq 1.0$, and $z > 1.0$. Therefore, instead of two offset parameters in the Hubble diagram as in \cite{Conley:2011ku}, we now have a total of six $\mathcal{M}$ parameters. The redshift extent that pairs of these parameters cover is illustrated in Fig.~\ref{fig:mass_step}, with the divisions centered on the fiducial model presented in \cite{Rigault:2013gux}. Clearly, once their amplitudes are allowed to float, these nuisance parameters will do a much better job recovering the redshift dependence of the mass step than a single pair of $\mathcal{M}$ parameters for the whole redshift range. We succeeded in marginalizing analytically over these six parameters with flat priors and including covariance between SNe with different $\mathcal{M}$ (see the Appendix).

The result is shown in Fig.~\ref{fig:six_M_Om_w}. In the left panel, the filled blue contours show the 68.3\%, 95.4\% and 99.7\% constraints for the usual case with two $\mathcal{M}$ parameters, while the open red contours show the result for the six $\mathcal{M}$ parameters. The constraints clearly weaken, although not as much as one might expect with four extra parameters introduced at the level of the Hubble diagram. This ``self-calibration'' serves to effectively protect against departures from the standard-candle assumption. Remarkably, when SN Ia data is combined with CMB and BAO data, the resulting constraints are barely weakened because the lengthening of the contours occurs mainly in the direction that is very well constrained by the complementary data sets. On the other hand, the right panel of Fig.~\ref{fig:six_M_Om_w} shows that the best-fit value of $w$ is shifted appreciably, illustrating the sensitivity of dark energy constraints to systematic effects in SN Ia measurements.

\subsection{Scanning through SN observables} \label{sec:biases}

Are SNe in any given redshift range of the SNLS3 compilation responsible for shifting the equation of state to phantom values? We examine this issue in Fig.~\ref{fig:residuals}, where we show the residuals in the Hubble diagram relative to $w = -1$ for SNe binned in $\Delta z = 0.075$ bins. In this analysis, we have assumed the same cosmology for both SNLS3 and Union2.1 ($\Omega_m = 0.3$, $w = -1$), where $\Omega_m$ is roughly the best-fit value from CMB + BAO. We fix the stretch, color, and $\mathcal{M}$ parameters at their best-fit values separately for each SN Ia data set. We see that the two data sets are consistent at the $1\sigma$ level in all bins except at $z \simeq 0.95$, where they are consistent at $2\sigma$. Therefore, the agreement between the two Hubble diagrams seems excellent.

\begin{figure*}[t]
\includegraphics[width=\textwidth]{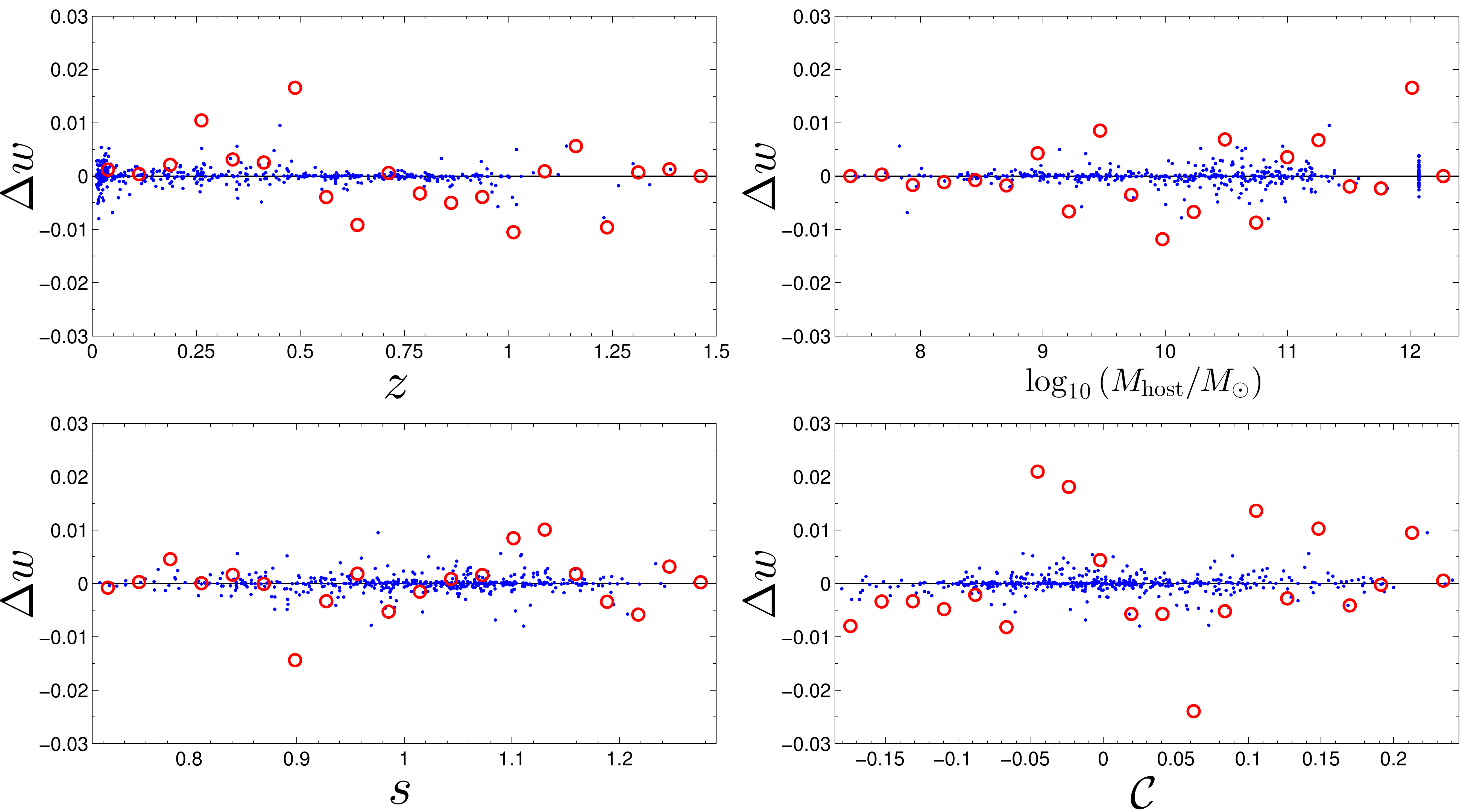}
\caption{Effect of each individual SNLS3 SN on the combined constraint on the equation of state, as a function of redshift (top left), host galaxy stellar mass (top right), stretch (bottom left), and color (bottom right). The blue points show the shift $\Delta w$ in the final constraint on $w$ due to each individual SN. The red circles show the combined (summed) pull from each bin in the particular quantity.}
\label{fig:biases}
\end{figure*}

We can get an even more accurate picture of the redshift dependence of SN constraints on the equation of state. The individual points in Fig.~\ref{fig:biases} show the effect of adding a single SN to the combination of Planck + BAO + SNLS3. For practical purposes,\footnote{The computed mean value depends on the precise likelihood ratios between different points in a grid of parameter values, but a simple numerical estimate of the maximum-likelihood value will only reflect changes that are of order the grid spacing or larger.} we compare the mean equation of state rather than the maximum of the likelihood. The red circles denote the total (summed) contribution of SNe per $\Delta z = 0.075$ bin. We see that no redshift bin contributes to a negative shift in the mean equation of state more than about 0.01.

The other three panels in Fig.~\ref{fig:biases} show the individual SNLS3 SN contributions to $w$ as a function of stretch, color, and host galaxy stellar mass. As before, the red circles denote the summed contribution of all SNe in a given bin in the quantity shown. As in the redshift scan, we do not observe any correlation or particular region in the stretch, color, or host-mass spaces that is chiefly responsible for shifting the equation of state.

\subsection{External \texorpdfstring{$H_0$}{H0} Prior} \label{sec:h0prior}

Adding a prior corresponding to an external measurement of the Hubble constant with a small error bar has an important effect on our results. This is easy to understand: Given that the CMB essentially pins down the physical matter density $\Omega_m h^2$ (for example, to better than 2\% with Planck), $\delta \ln \Omega_m \simeq -2 \delta \ln h$ and therefore a higher value of $H_0$ corresponds to a lower value of $\Omega_m$. For a lower $\Omega_m$, the degenerate direction of CMB + BAO constraints leads to a more negative $w$. Therefore, we would expect that higher values of $H_0$ lead to more negative $w$, and vice versa.

\begin{figure*}[t]
\includegraphics[width=0.85\textwidth]{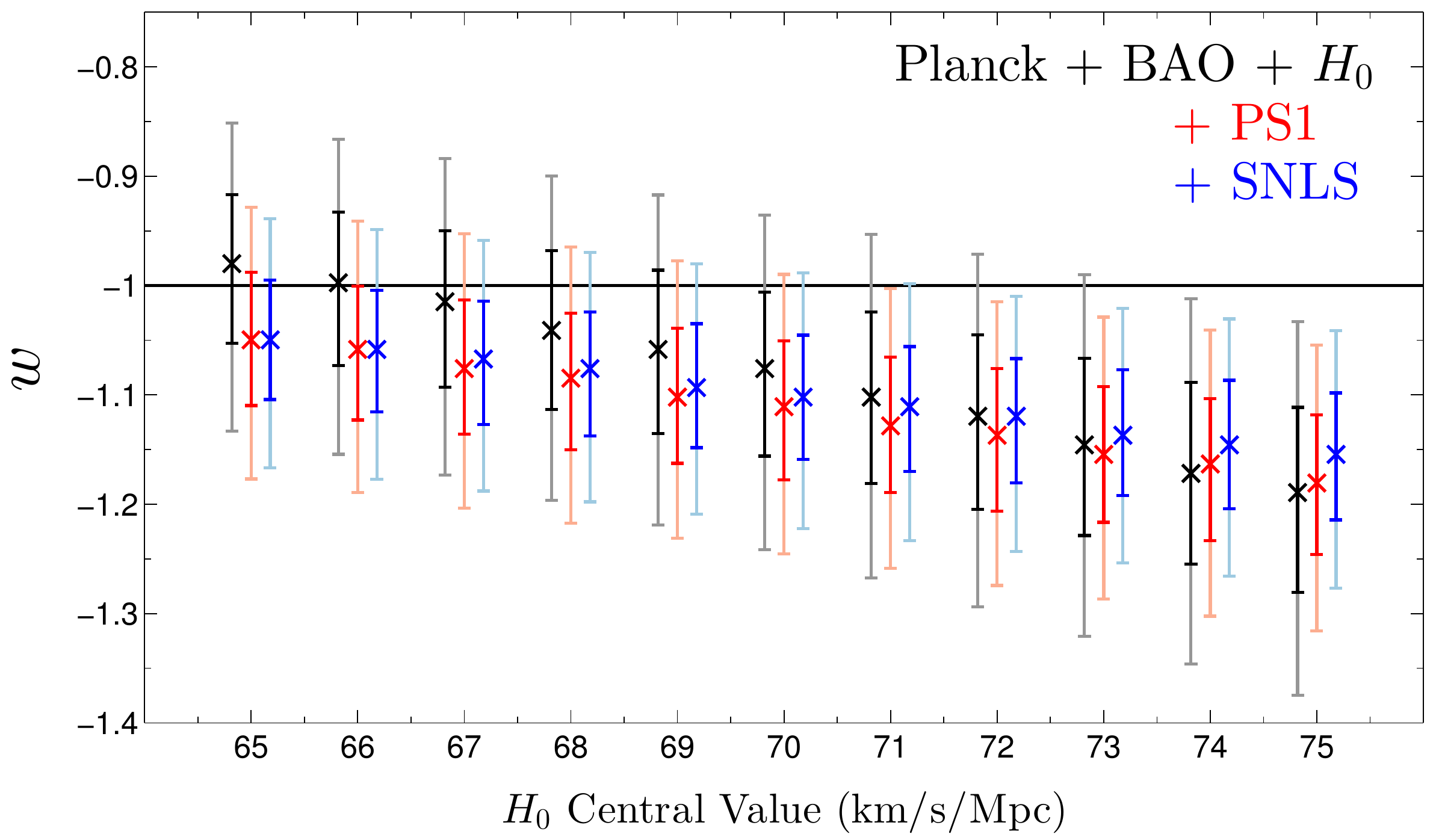}
\caption{Effect of an external $H_0$ prior on the constant equation of state. We show the effect on Planck + BAO constraints (black) and on combined Planck + BAO + SN constraints separately for PS1 (red) and SNLS3 (blue), where the error bars bound 68.3\% and 95.4\% of the likelihood for $w$. The external prior has an uncertainty of $2.4~\text{km/s/Mpc}$ in each case, mimicking the uncertainty in the \citet{Riess:2011yx} measurement.}
\label{fig:H0}
\end{figure*}

This expectation is confirmed by our explicit tests with the current data, shown in Fig.~\ref{fig:H0}. Here we use the same Hubble constant measurement error of $\pm 2.4~\text{km/s/Mpc}$ as reported by \citet{Riess:2011yx}, but instead of adopting the central value of 73.8~km/s/Mpc, we vary the central value as an integer in the range $H_0 \in [65 , 75]~\text{km/s/Mpc}$, one value at a time. We show the final constraint on the dark energy equation of state using the CMB + BAO + $H_0$ data, with or without the addition of the PS1 or SNLS3 SN data, as a function of the $H_0$ central value. For the external prior $H_0 = 74 \pm 2.4~\text{km/s/Mpc}$, we recover results similar to \cite{Rest:2013bya} that favor $w < -1$ at $\sim 2.5\sigma$. However, for a smaller central value of $H_0$ ($\leq 70~\text{km/s/Mpc}$), the results are consistent with $w = -1$ at the $2\sigma$ level or less, and for an even smaller central value ($\simeq 66~\text{km/s/Mpc}$), we find the results consistent with $w = -1$ at $1\sigma$.

\section{Conclusions} \label{sec:conclude}

We studied geometric constraints on the dark energy equation of state from recent SN Ia data complemented with distance measurements from the CMB and a compilation of BAO results. For the SNLS3 and PS1 SN data sets, the combined SN Ia + BAO + Planck data favor a phantom equation of state where $w < -1$ at $\sim 1.9\sigma$ confidence (see Fig.~\ref{fig:wlike}), in good agreement with the corresponding results reported in the original SN Ia and Planck papers. Evidence for a phantom equation of state is weaker if WMAP9 is used instead of Planck, while the Union2.1 data set is consistent with the cosmological-constant value $w = -1$ when combined with either CMB data set.

We have tested for a possible presence of systematics correlated with SN properties -- their redshifts, stretch factors, colors, and host galaxy masses. We find no evidence of a trend or that a particular range of any of these properties contributes to pushing $w < -1$; rather, the hints of a phantom equation of state appear to be uncorrelated with these basic SN observables. We have also investigated the effect of modeling the redshift dependence of the host galaxy mass step of SN luminosities, assigning up to six separate $\mathcal{M}$ parameters for different mass and redshift bins. The additional nuisance parameters shift the SN Ia constraints sufficiently that, when combined with BAO and CMB data, they allow agreement with $w = -1$ at $\sim 1\sigma$. Therefore, a more generous allowance for the temporal evolution of the dependence of SN luminosity on host galaxy mass removes the evidence for $w < -1$. The hope for the future is that independent observations can pin down the environmental dependence of SN luminosities and make it possible to account for such subtle but important systematic effects in a consistent way and without the damaging effect of extra nuisance parameters at the level of the Hubble diagram.

External measurements of the Hubble constant play a particularly important role in the final constraints on $w$. This is shown in Fig.~\ref{fig:H0}, where we illustrate the effect of adding a measurement of $H_0$ with an error of 2.4~km/s/Mpc as in \citet{Riess:2011yx}, but with the central value varied from 65-75~km/s/Mpc. Clearly, interesting $> 2\sigma$ evidence for the phantom equation of state is present only when the central value is somewhat large: $H_0 \gtrsim 71~\text{km/s/Mpc}$. Therefore, as first clearly argued by \citet{Hu:2004kn}, with excellent CMB constraints the Hubble constant measurements and their interpretation (e.g.\ \cite{Efstathiou:2013via, Marra:2013rba}) are among the most important inputs in determining the dark energy equation of state.

Although we have taken Planck + BAO data at face value throughout most of our analysis, it is worth mentioning that systematics may be present in these data as well. This is particularly true for the Planck data, given that its analysis is still in the early stages and given the moderate tension between Planck results and both WMAP and growth measurements (e.g.\ \cite{Macaulay:2013swa}). Indeed, a recent re-analysis of Planck data \cite{Spergel:2013rxa} resulted in parameter shifts that somewhat reduce the tension with WMAP, and substantial ongoing work is focused on understanding systematics in growth measurements.

With current data we therefore find ourselves in an interesting situation in which we can make the following statement at $2\sigma$ confidence: Given Planck data, either the SNLS3 and PS1 data have systematics that remain unaccounted for or the Hubble constant is below 71~km/s/Mpc; else the dark energy equation of state is indeed phantom.

\begin{acknowledgments}

We thank Chris Blake, Alex Conley, Eric Linder, and Adam Riess for thoughtful comments on the manuscript and useful discussions. We also thank all three supernova teams for making their data and covariance matrices publicly available. Our work has been supported by DOE Grant No.\ DE-SC0007859 and NSF Grant No.\ AST 1210974.

\end{acknowledgments}

\appendix*
\section{Analytic marginalization over multiple \texorpdfstring{$\mathcal{M}$}{M}}

The computational cost of varying additional nuisance parameters in a likelihood analysis can be reduced by marginalizing analytically over some of these parameters. The expression for the marginalized $\chi^2$ will be more complicated, but fewer likelihoods will need to be computed in the analysis (in a brute-force grid search, many times fewer). Here we extend Appendix~C of \cite{Conley:2011ku} and outline a procedure to marginalize analytically over a model with more than two $\mathcal{M}$.

The analytic marginalization over the single $\mathcal{M}$ in Eq.~\eqref{eq:mth} is very straightforward. In this case, the marginalized $\chi^2$ (for a flat prior) is given by
\begin{align}
\chi^2_\text{marg} &= -2 \log \int_{-\infty}^{\infty} e^{-\chi^2/2} \; \text{d}\mathcal{M} \nonumber \\
&= X_{00} + \log \left(\frac{X_{11}}{2 \pi} \right) - \frac{X_{10}^2}{X_{11}} \ ,
\end{align}
where the unmarginalized $\chi^2$ is given by Eq.~\eqref{eq:covchi2} and
\begin{align*}
X_{00} &= \bm{\delta}^\intercal \, \mathbf{C}^{-1} \bm{\delta} \\
X_{10} &= \bm{1}^\intercal \, \mathbf{C}^{-1} \bm{\delta} \\
X_{11} &= \bm{1}^\intercal \, \mathbf{C}^{-1} \bm{1} \ .
\end{align*}
In the above, $\bm{1}$ is a vector of ones and $\bm{\delta}$ is the vector of magnitude residuals without the $\mathcal{M}$ term:
\begin{equation*}
\bm{\delta} = \Delta \mathbf{m} + \mathcal{M} \, \bm{1} = \mathbf{m}_\text{corr} - 5 \log_{10} \left(\frac{H_0}{c} \, \mathbf{D_L} \right) \ .
\end{equation*}

Marginalizing over a two-$\mathcal{M}$ model is a bit more complicated due to covariance between SNe with different $\mathcal{M}$. Ignoring this covariance will make things much simpler, but the resulting constraints will be biased and the effect can be significant. For the two-$\mathcal{M}$ case, we modify Eq.~\eqref{eq:mth} so that the vector of predicted SN magnitudes is given by
\begin{equation*}
\mathbf{m}_\text{th} = 5 \log_{10} \left(\frac{H_0}{c} \, \mathbf{D_L} \right) + \mathcal{M}_1 \, \mathbf{x}_1 + \mathcal{M}_2 \, \mathbf{x}_2 \ ,
\end{equation*}
where $\mathbf{x}_1$ ($\mathbf{x}_2$) is a vector with ones for SNe described by $\mathcal{M}_1$ ($\mathcal{M}_2$), and zeros otherwise. The marginalized $\chi^2$ is given by
\begin{align}
\chi^2_\text{marg} &= X_{00} + \log \left(\frac{X}{4 \pi^2} \right) \\
&- \frac{1}{X} \left(X_{10}^2 X_{22} + X_{20}^2 X_{11} - 2 X_{10} X_{20} X_{12} \right) \ , \nonumber
\end{align}
where
\begin{align*}
X_{00} &= \bm{\delta}^\intercal \, \mathbf{C}^{-1} \bm{\delta} \\
X_{10} &= \mathbf{x}_1^\intercal \, \mathbf{C}^{-1} \bm{\delta} \\
X_{20} &= \mathbf{x}_2^\intercal \, \mathbf{C}^{-1} \bm{\delta} \\
X_{11} &= \mathbf{x}_1^\intercal \, \mathbf{C}^{-1} \mathbf{x}_1 \\
X_{22} &= \mathbf{x}_2^\intercal \, \mathbf{C}^{-1} \mathbf{x}_2 \\
X_{12} &= \mathbf{x}_1^\intercal \, \mathbf{C}^{-1} \mathbf{x}_2 \\
X &= X_{11} X_{22} - X_{12}^2 \ .
\end{align*}

One can extrapolate the two-$\mathcal{M}$ case above to cases with three or more $\mathcal{M}$. This is straightforward, but the result quickly becomes very messy. Using the familiar result for the integral of a Gaussian function, one can compute the expression for several $\mathcal{M}$ using algebraic manipulation software. Explicit Mathematica code for marginalizing over six $\mathcal{M}$ (as in this work) is given below. The result must be simplified at intermediate steps to avoid long computation times, memory problems, or unmanageable final expressions. Still, the computation below takes several minutes, and the resulting expressions are very long (several pages of small type). This example can easily be modified for more or fewer $\mathcal{M}$ and for slightly different parametrizations of the SN magnitude (for instance, treating one $\mathcal{M}$ as a $\Delta \mathcal{M}$ so that some SNe are described by more than one such nuisance parameter).

\clearpage

\begin{widetext}
\begin{small}
\begin{verbatim}
fexp[{a_,b_,c_}] := b^2/(4*a)+c
fpre[{a_,b_,c_}] := a/Pi
g[f_,x_] := {-1*Coefficient[f,x^2],Coefficient[f,x],f/.x->0}
pre := 1
exp := -1/2*(X00-2*M1*X10-2*M2*X20-2*M3*X30-2*M4*X40-2*M5*X50-2*M6*X60
+M1^2*X11+M2^2*X22+M3^2*X33+M4^2*X44+M5^2*X55+M6^2*X66
+2*M1*M2*X12+2*M1*M3*X13+2*M1*M4*X14+2*M1*M5*X15+2*M1*M6*X16+2*M2*M3*X23+2*M2*M4*X24+2*M2*M5*X25
+2*M2*M6*X26+2*M3*M4*X34+2*M3*M5*X35+2*M3*M6*X36+2*M4*M5*X45+2*M4*M6*X46+2*M5*M6*X56)
Do[{
coeffs = g[exp,Mi];
pre = FullSimplify[Expand[pre*fpre[coeffs]]];
exp = FullSimplify[Expand[fexp[coeffs]]]},{Mi,{M1,M2,M3,M4,M5,M6}}]
chisq_marg := Log[pre]-2*exp
\end{verbatim}
\end{small}
\end{widetext}

\bibliography{const_w}

\begin{thebibliography}{55}%
\makeatletter
\providecommand \@ifxundefined [1]{%
 \@ifx{#1\undefined}
}%
\providecommand \@ifnum [1]{%
 \ifnum #1\expandafter \@firstoftwo
 \else \expandafter \@secondoftwo
 \fi
}%
\providecommand \@ifx [1]{%
 \ifx #1\expandafter \@firstoftwo
 \else \expandafter \@secondoftwo
 \fi
}%
\providecommand \natexlab [1]{#1}%
\providecommand \enquote  [1]{``#1''}%
\providecommand \bibnamefont  [1]{#1}%
\providecommand \bibfnamefont [1]{#1}%
\providecommand \citenamefont [1]{#1}%
\providecommand \href@noop [0]{\@secondoftwo}%
\providecommand \href [0]{\begingroup \@sanitize@url \@href}%
\providecommand \@href[1]{\@@startlink{#1}\@@href}%
\providecommand \@@href[1]{\endgroup#1\@@endlink}%
\providecommand \@sanitize@url [0]{\catcode `\\12\catcode `\$12\catcode
  `\&12\catcode `\#12\catcode `\^12\catcode `\_12\catcode `\%12\relax}%
\providecommand \@@startlink[1]{}%
\providecommand \@@endlink[0]{}%
\providecommand \url  [0]{\begingroup\@sanitize@url \@url }%
\providecommand \@url [1]{\endgroup\@href {#1}{\urlprefix }}%
\providecommand \urlprefix  [0]{URL }%
\providecommand \Eprint [0]{\href }%
\providecommand \doibase [0]{http://dx.doi.org/}%
\providecommand \selectlanguage [0]{\@gobble}%
\providecommand \bibinfo  [0]{\@secondoftwo}%
\providecommand \bibfield  [0]{\@secondoftwo}%
\providecommand \translation [1]{[#1]}%
\providecommand \BibitemOpen [0]{}%
\providecommand \bibitemStop [0]{}%
\providecommand \bibitemNoStop [0]{.\EOS\space}%
\providecommand \EOS [0]{\spacefactor3000\relax}%
\providecommand \BibitemShut  [1]{\csname bibitem#1\endcsname}%
\let\auto@bib@innerbib\@empty
\bibitem [{\citenamefont {Davis}\ \emph {et~al.}(2007)\citenamefont {Davis},
  \citenamefont {Mortsell}, \citenamefont {Sollerman}, \citenamefont {Becker},
  \citenamefont {Blondin} \emph {et~al.}}]{Davis:2007na}%
  \BibitemOpen
  \bibfield  {author} {\bibinfo {author} {\bibfnamefont {T.~M.}\ \bibnamefont
  {Davis}}, \bibinfo {author} {\bibfnamefont {E.}~\bibnamefont {Mortsell}},
  \bibinfo {author} {\bibfnamefont {J.}~\bibnamefont {Sollerman}}, \bibinfo
  {author} {\bibfnamefont {A.}~\bibnamefont {Becker}}, \bibinfo {author}
  {\bibfnamefont {S.}~\bibnamefont {Blondin}},  \emph {et~al.},\ }\href
  {\doibase 10.1086/519988} {\bibfield  {journal} {\bibinfo  {journal}
  {Astrophys.J.}\ }\textbf {\bibinfo {volume} {666}},\ \bibinfo {pages} {716}
  (\bibinfo {year} {2007})},\ \Eprint {http://arxiv.org/abs/astro-ph/0701510}
  {arXiv:astro-ph/0701510 [astro-ph]} \BibitemShut {NoStop}%
\bibitem [{\citenamefont {Riess}\ \emph {et~al.}(2007)\citenamefont {Riess},
  \citenamefont {Strolger}, \citenamefont {Casertano}, \citenamefont
  {Ferguson}, \citenamefont {Mobasher} \emph {et~al.}}]{Riess:2006fw}%
  \BibitemOpen
  \bibfield  {author} {\bibinfo {author} {\bibfnamefont {A.~G.}\ \bibnamefont
  {Riess}}, \bibinfo {author} {\bibfnamefont {L.-G.}\ \bibnamefont {Strolger}},
  \bibinfo {author} {\bibfnamefont {S.}~\bibnamefont {Casertano}}, \bibinfo
  {author} {\bibfnamefont {H.~C.}\ \bibnamefont {Ferguson}}, \bibinfo {author}
  {\bibfnamefont {B.}~\bibnamefont {Mobasher}},  \emph {et~al.},\ }\href
  {\doibase 10.1086/510378} {\bibfield  {journal} {\bibinfo  {journal}
  {Astrophys.J.}\ }\textbf {\bibinfo {volume} {659}},\ \bibinfo {pages} {98}
  (\bibinfo {year} {2007})},\ \Eprint {http://arxiv.org/abs/astro-ph/0611572}
  {arXiv:astro-ph/0611572 [astro-ph]} \BibitemShut {NoStop}%
\bibitem [{\citenamefont {Knop}\ \emph {et~al.}(2003)\citenamefont {Knop} \emph
  {et~al.}}]{Knop:2003iy}%
  \BibitemOpen
  \bibfield  {author} {\bibinfo {author} {\bibfnamefont {R.~A.}\ \bibnamefont
  {Knop}} \emph {et~al.} (\bibinfo {collaboration} {Supernova Cosmology
  Project}),\ }\href {\doibase 10.1086/378560} {\bibfield  {journal} {\bibinfo
  {journal} {Astrophys.J.}\ }\textbf {\bibinfo {volume} {598}},\ \bibinfo
  {pages} {102} (\bibinfo {year} {2003})},\ \Eprint
  {http://arxiv.org/abs/astro-ph/0309368} {arXiv:astro-ph/0309368 [astro-ph]}
  \BibitemShut {NoStop}%
\bibitem [{\citenamefont {Astier}\ \emph {et~al.}(2006)\citenamefont {Astier}
  \emph {et~al.}}]{Astier:2005qq}%
  \BibitemOpen
  \bibfield  {author} {\bibinfo {author} {\bibfnamefont {P.}~\bibnamefont
  {Astier}} \emph {et~al.} (\bibinfo {collaboration} {SNLS Collaboration}),\
  }\href {\doibase 10.1051/0004-6361:20054185} {\bibfield  {journal} {\bibinfo
  {journal} {Astron.Astrophys.}\ }\textbf {\bibinfo {volume} {447}},\ \bibinfo
  {pages} {31} (\bibinfo {year} {2006})},\ \Eprint
  {http://arxiv.org/abs/astro-ph/0510447} {arXiv:astro-ph/0510447 [astro-ph]}
  \BibitemShut {NoStop}%
\bibitem [{\citenamefont {Wood-Vasey}\ \emph {et~al.}(2007)\citenamefont
  {Wood-Vasey} \emph {et~al.}}]{WoodVasey:2007jb}%
  \BibitemOpen
  \bibfield  {author} {\bibinfo {author} {\bibfnamefont {W.~M.}\ \bibnamefont
  {Wood-Vasey}} \emph {et~al.} (\bibinfo {collaboration} {ESSENCE
  Collaboration}),\ }\href {\doibase 10.1086/518642} {\bibfield  {journal}
  {\bibinfo  {journal} {Astrophys.J.}\ }\textbf {\bibinfo {volume} {666}},\
  \bibinfo {pages} {694} (\bibinfo {year} {2007})},\ \Eprint
  {http://arxiv.org/abs/astro-ph/0701041} {arXiv:astro-ph/0701041 [astro-ph]}
  \BibitemShut {NoStop}%
\bibitem [{\citenamefont {Rubin}\ \emph {et~al.}(2009)\citenamefont {Rubin},
  \citenamefont {Linder}, \citenamefont {Kowalski}, \citenamefont {Aldering},
  \citenamefont {Amanullah} \emph {et~al.}}]{Rubin:2008wq}%
  \BibitemOpen
  \bibfield  {author} {\bibinfo {author} {\bibfnamefont {D.}~\bibnamefont
  {Rubin}}, \bibinfo {author} {\bibfnamefont {E.}~\bibnamefont {Linder}},
  \bibinfo {author} {\bibfnamefont {M.}~\bibnamefont {Kowalski}}, \bibinfo
  {author} {\bibfnamefont {G.}~\bibnamefont {Aldering}}, \bibinfo {author}
  {\bibfnamefont {R.}~\bibnamefont {Amanullah}},  \emph {et~al.},\ }\href
  {\doibase 10.1088/0004-637X/695/1/391} {\bibfield  {journal} {\bibinfo
  {journal} {Astrophys.J.}\ }\textbf {\bibinfo {volume} {695}},\ \bibinfo
  {pages} {391} (\bibinfo {year} {2009})},\ \Eprint
  {http://arxiv.org/abs/0807.1108} {arXiv:0807.1108 [astro-ph]} \BibitemShut
  {NoStop}%
\bibitem [{\citenamefont {Freedman}\ \emph {et~al.}(2009)\citenamefont
  {Freedman}, \citenamefont {Burns}, \citenamefont {Phillips}, \citenamefont
  {Wyatt}, \citenamefont {Persson} \emph {et~al.}}]{Freedman:2009vv}%
  \BibitemOpen
  \bibfield  {author} {\bibinfo {author} {\bibfnamefont {W.~L.}\ \bibnamefont
  {Freedman}}, \bibinfo {author} {\bibfnamefont {C.~R.}\ \bibnamefont {Burns}},
  \bibinfo {author} {\bibfnamefont {M.}~\bibnamefont {Phillips}}, \bibinfo
  {author} {\bibfnamefont {P.}~\bibnamefont {Wyatt}}, \bibinfo {author}
  {\bibfnamefont {S.}~\bibnamefont {Persson}},  \emph {et~al.},\ }\href
  {\doibase 10.1088/0004-637X/704/2/1036} {\bibfield  {journal} {\bibinfo
  {journal} {Astrophys.J.}\ }\textbf {\bibinfo {volume} {704}},\ \bibinfo
  {pages} {1036} (\bibinfo {year} {2009})},\ \Eprint
  {http://arxiv.org/abs/0907.4524} {arXiv:0907.4524 [astro-ph.CO]} \BibitemShut
  {NoStop}%
\bibitem [{\citenamefont {Kowalski}\ \emph {et~al.}(2008)\citenamefont
  {Kowalski} \emph {et~al.}}]{Kowalski:2008ez}%
  \BibitemOpen
  \bibfield  {author} {\bibinfo {author} {\bibfnamefont {M.}~\bibnamefont
  {Kowalski}} \emph {et~al.} (\bibinfo {collaboration} {Supernova Cosmology
  Project}),\ }\href {\doibase 10.1086/589937} {\bibfield  {journal} {\bibinfo
  {journal} {Astrophys.J.}\ }\textbf {\bibinfo {volume} {686}},\ \bibinfo
  {pages} {749} (\bibinfo {year} {2008})},\ \Eprint
  {http://arxiv.org/abs/0804.4142} {arXiv:0804.4142 [astro-ph]} \BibitemShut
  {NoStop}%
\bibitem [{\citenamefont {Hicken}\ \emph {et~al.}(2009)\citenamefont {Hicken},
  \citenamefont {Wood-Vasey}, \citenamefont {Blondin}, \citenamefont {Challis},
  \citenamefont {Jha} \emph {et~al.}}]{Hicken:2009dk}%
  \BibitemOpen
  \bibfield  {author} {\bibinfo {author} {\bibfnamefont {M.}~\bibnamefont
  {Hicken}}, \bibinfo {author} {\bibfnamefont {W.~M.}\ \bibnamefont
  {Wood-Vasey}}, \bibinfo {author} {\bibfnamefont {S.}~\bibnamefont {Blondin}},
  \bibinfo {author} {\bibfnamefont {P.}~\bibnamefont {Challis}}, \bibinfo
  {author} {\bibfnamefont {S.}~\bibnamefont {Jha}},  \emph {et~al.},\ }\href
  {\doibase 10.1088/0004-637X/700/2/1097} {\bibfield  {journal} {\bibinfo
  {journal} {Astrophys.J.}\ }\textbf {\bibinfo {volume} {700}},\ \bibinfo
  {pages} {1097} (\bibinfo {year} {2009})},\ \Eprint
  {http://arxiv.org/abs/0901.4804} {arXiv:0901.4804 [astro-ph.CO]} \BibitemShut
  {NoStop}%
\bibitem [{\citenamefont {Kessler}\ \emph {et~al.}(2009)\citenamefont
  {Kessler}, \citenamefont {Becker}, \citenamefont {Cinabro}, \citenamefont
  {Vanderplas}, \citenamefont {Frieman} \emph {et~al.}}]{Kessler:2009ys}%
  \BibitemOpen
  \bibfield  {author} {\bibinfo {author} {\bibfnamefont {R.}~\bibnamefont
  {Kessler}}, \bibinfo {author} {\bibfnamefont {A.}~\bibnamefont {Becker}},
  \bibinfo {author} {\bibfnamefont {D.}~\bibnamefont {Cinabro}}, \bibinfo
  {author} {\bibfnamefont {J.}~\bibnamefont {Vanderplas}}, \bibinfo {author}
  {\bibfnamefont {J.~A.}\ \bibnamefont {Frieman}},  \emph {et~al.},\ }\href
  {\doibase 10.1088/0067-0049/185/1/32} {\bibfield  {journal} {\bibinfo
  {journal} {Astrophys.J.Suppl.}\ }\textbf {\bibinfo {volume} {185}},\ \bibinfo
  {pages} {32} (\bibinfo {year} {2009})},\ \Eprint
  {http://arxiv.org/abs/0908.4274} {arXiv:0908.4274 [astro-ph.CO]} \BibitemShut
  {NoStop}%
\bibitem [{\citenamefont {Amanullah}\ \emph {et~al.}(2010)\citenamefont
  {Amanullah}, \citenamefont {Lidman}, \citenamefont {Rubin}, \citenamefont
  {Aldering}, \citenamefont {Astier} \emph {et~al.}}]{Amanullah:2010vv}%
  \BibitemOpen
  \bibfield  {author} {\bibinfo {author} {\bibfnamefont {R.}~\bibnamefont
  {Amanullah}}, \bibinfo {author} {\bibfnamefont {C.}~\bibnamefont {Lidman}},
  \bibinfo {author} {\bibfnamefont {D.}~\bibnamefont {Rubin}}, \bibinfo
  {author} {\bibfnamefont {G.}~\bibnamefont {Aldering}}, \bibinfo {author}
  {\bibfnamefont {P.}~\bibnamefont {Astier}},  \emph {et~al.},\ }\href
  {\doibase 10.1088/0004-637X/716/1/712} {\bibfield  {journal} {\bibinfo
  {journal} {Astrophys.J.}\ }\textbf {\bibinfo {volume} {716}},\ \bibinfo
  {pages} {712} (\bibinfo {year} {2010})},\ \Eprint
  {http://arxiv.org/abs/1004.1711} {arXiv:1004.1711 [astro-ph.CO]} \BibitemShut
  {NoStop}%
\bibitem [{\citenamefont {Suzuki}\ \emph {et~al.}(2012)\citenamefont {Suzuki},
  \citenamefont {Rubin}, \citenamefont {Lidman}, \citenamefont {Aldering},
  \citenamefont {Amanullah} \emph {et~al.}}]{Suzuki:2011hu}%
  \BibitemOpen
  \bibfield  {author} {\bibinfo {author} {\bibfnamefont {N.}~\bibnamefont
  {Suzuki}}, \bibinfo {author} {\bibfnamefont {D.}~\bibnamefont {Rubin}},
  \bibinfo {author} {\bibfnamefont {C.}~\bibnamefont {Lidman}}, \bibinfo
  {author} {\bibfnamefont {G.}~\bibnamefont {Aldering}}, \bibinfo {author}
  {\bibfnamefont {R.}~\bibnamefont {Amanullah}},  \emph {et~al.},\ }\href
  {\doibase 10.1088/0004-637X/746/1/85} {\bibfield  {journal} {\bibinfo
  {journal} {Astrophys.J.}\ }\textbf {\bibinfo {volume} {746}},\ \bibinfo
  {pages} {85} (\bibinfo {year} {2012})},\ \Eprint
  {http://arxiv.org/abs/1105.3470} {arXiv:1105.3470 [astro-ph.CO]} \BibitemShut
  {NoStop}%
\bibitem [{\citenamefont {Huterer}\ and\ \citenamefont
  {Cooray}(2005)}]{Huterer:2004ch}%
  \BibitemOpen
  \bibfield  {author} {\bibinfo {author} {\bibfnamefont {D.}~\bibnamefont
  {Huterer}}\ and\ \bibinfo {author} {\bibfnamefont {A.}~\bibnamefont
  {Cooray}},\ }\href {\doibase 10.1103/PhysRevD.71.023506} {\bibfield
  {journal} {\bibinfo  {journal} {Phys.Rev.}\ }\textbf {\bibinfo {volume}
  {D71}},\ \bibinfo {pages} {023506} (\bibinfo {year} {2005})},\ \Eprint
  {http://arxiv.org/abs/astro-ph/0404062} {arXiv:astro-ph/0404062 [astro-ph]}
  \BibitemShut {NoStop}%
\bibitem [{\citenamefont {Alam}\ \emph {et~al.}(2007)\citenamefont {Alam},
  \citenamefont {Sahni},\ and\ \citenamefont {Starobinsky}}]{Alam:2006kj}%
  \BibitemOpen
  \bibfield  {author} {\bibinfo {author} {\bibfnamefont {U.}~\bibnamefont
  {Alam}}, \bibinfo {author} {\bibfnamefont {V.}~\bibnamefont {Sahni}}, \ and\
  \bibinfo {author} {\bibfnamefont {A.~A.}\ \bibnamefont {Starobinsky}},\
  }\href {\doibase 10.1088/1475-7516/2007/02/011} {\bibfield  {journal}
  {\bibinfo  {journal} {JCAP}\ }\textbf {\bibinfo {volume} {0702}},\ \bibinfo
  {pages} {011} (\bibinfo {year} {2007})},\ \Eprint
  {http://arxiv.org/abs/astro-ph/0612381} {arXiv:astro-ph/0612381 [astro-ph]}
  \BibitemShut {NoStop}%
\bibitem [{\citenamefont {Nesseris}\ and\ \citenamefont
  {Perivolaropoulos}(2007)}]{Nesseris:2006ey}%
  \BibitemOpen
  \bibfield  {author} {\bibinfo {author} {\bibfnamefont {S.}~\bibnamefont
  {Nesseris}}\ and\ \bibinfo {author} {\bibfnamefont {L.}~\bibnamefont
  {Perivolaropoulos}},\ }\href {\doibase 10.1088/1475-7516/2007/02/025}
  {\bibfield  {journal} {\bibinfo  {journal} {JCAP}\ }\textbf {\bibinfo
  {volume} {0702}},\ \bibinfo {pages} {025} (\bibinfo {year} {2007})},\ \Eprint
  {http://arxiv.org/abs/astro-ph/0612653} {arXiv:astro-ph/0612653 [astro-ph]}
  \BibitemShut {NoStop}%
\bibitem [{\citenamefont {Zhao}\ and\ \citenamefont
  {Zhang}(2010)}]{Zhao:2009ti}%
  \BibitemOpen
  \bibfield  {author} {\bibinfo {author} {\bibfnamefont {G.-B.}\ \bibnamefont
  {Zhao}}\ and\ \bibinfo {author} {\bibfnamefont {X.-m.}\ \bibnamefont
  {Zhang}},\ }\href {\doibase 10.1103/PhysRevD.81.043518} {\bibfield  {journal}
  {\bibinfo  {journal} {Phys.Rev.}\ }\textbf {\bibinfo {volume} {D81}},\
  \bibinfo {pages} {043518} (\bibinfo {year} {2010})},\ \Eprint
  {http://arxiv.org/abs/0908.1568} {arXiv:0908.1568 [astro-ph.CO]} \BibitemShut
  {NoStop}%
\bibitem [{\citenamefont {Ade}\ \emph {et~al.}(2013)\citenamefont {Ade} \emph
  {et~al.}}]{Ade:2013zuv}%
  \BibitemOpen
  \bibfield  {author} {\bibinfo {author} {\bibfnamefont {P.}~\bibnamefont
  {Ade}} \emph {et~al.} (\bibinfo {collaboration} {Planck Collaboration}),\
  }\href@noop {} {\  (\bibinfo {year} {2013})},\ \Eprint
  {http://arxiv.org/abs/1303.5076} {arXiv:1303.5076 [astro-ph.CO]} \BibitemShut
  {NoStop}%
\bibitem [{\citenamefont {Rest}\ \emph {et~al.}(2013)\citenamefont {Rest},
  \citenamefont {Scolnic}, \citenamefont {Foley}, \citenamefont {Huber},
  \citenamefont {Chornock} \emph {et~al.}}]{Rest:2013bya}%
  \BibitemOpen
  \bibfield  {author} {\bibinfo {author} {\bibfnamefont {A.}~\bibnamefont
  {Rest}}, \bibinfo {author} {\bibfnamefont {D.}~\bibnamefont {Scolnic}},
  \bibinfo {author} {\bibfnamefont {R.}~\bibnamefont {Foley}}, \bibinfo
  {author} {\bibfnamefont {M.}~\bibnamefont {Huber}}, \bibinfo {author}
  {\bibfnamefont {R.}~\bibnamefont {Chornock}},  \emph {et~al.},\ }\href@noop
  {} {\  (\bibinfo {year} {2013})},\ \Eprint {http://arxiv.org/abs/1310.3828}
  {arXiv:1310.3828 [astro-ph.CO]} \BibitemShut {NoStop}%
\bibitem [{\citenamefont {Scolnic}\ \emph {et~al.}(2013)\citenamefont
  {Scolnic}, \citenamefont {Rest}, \citenamefont {Riess}, \citenamefont
  {Huber}, \citenamefont {Foley} \emph {et~al.}}]{Scolnic:2013aya}%
  \BibitemOpen
  \bibfield  {author} {\bibinfo {author} {\bibfnamefont {D.}~\bibnamefont
  {Scolnic}}, \bibinfo {author} {\bibfnamefont {A.}~\bibnamefont {Rest}},
  \bibinfo {author} {\bibfnamefont {A.}~\bibnamefont {Riess}}, \bibinfo
  {author} {\bibfnamefont {M.}~\bibnamefont {Huber}}, \bibinfo {author}
  {\bibfnamefont {R.}~\bibnamefont {Foley}},  \emph {et~al.},\ }\href@noop {}
  {\  (\bibinfo {year} {2013})},\ \Eprint {http://arxiv.org/abs/1310.3824}
  {arXiv:1310.3824 [astro-ph.CO]} \BibitemShut {NoStop}%
\bibitem [{\citenamefont {Cheng}\ and\ \citenamefont
  {Huang}(2014)}]{Cheng:2013csa}%
  \BibitemOpen
  \bibfield  {author} {\bibinfo {author} {\bibfnamefont {C.}~\bibnamefont
  {Cheng}}\ and\ \bibinfo {author} {\bibfnamefont {Q.-G.}\ \bibnamefont
  {Huang}},\ }\href@noop {} {\bibfield  {journal} {\bibinfo  {journal}
  {Phys.Rev.}\ }\textbf {\bibinfo {volume} {D89}},\ \bibinfo {pages} {043003}
  (\bibinfo {year} {2014})},\ \Eprint {http://arxiv.org/abs/1306.4091}
  {arXiv:1306.4091 [astro-ph.CO]} \BibitemShut {NoStop}%
\bibitem [{\citenamefont {Xia}\ \emph {et~al.}(2013)\citenamefont {Xia},
  \citenamefont {Li},\ and\ \citenamefont {Zhang}}]{Xia:2013dea}%
  \BibitemOpen
  \bibfield  {author} {\bibinfo {author} {\bibfnamefont {J.-Q.}\ \bibnamefont
  {Xia}}, \bibinfo {author} {\bibfnamefont {H.}~\bibnamefont {Li}}, \ and\
  \bibinfo {author} {\bibfnamefont {X.}~\bibnamefont {Zhang}},\ }\href
  {\doibase 10.1103/PhysRevD.88.063501} {\bibfield  {journal} {\bibinfo
  {journal} {Phys.Rev.}\ }\textbf {\bibinfo {volume} {D88}},\ \bibinfo {pages}
  {063501} (\bibinfo {year} {2013})},\ \Eprint {http://arxiv.org/abs/1308.0188}
  {arXiv:1308.0188 [astro-ph.CO]} \BibitemShut {NoStop}%
\bibitem [{\citenamefont {Frieman}\ \emph {et~al.}(2008)\citenamefont
  {Frieman}, \citenamefont {Turner},\ and\ \citenamefont
  {Huterer}}]{Frieman:2008sn}%
  \BibitemOpen
  \bibfield  {author} {\bibinfo {author} {\bibfnamefont {J.}~\bibnamefont
  {Frieman}}, \bibinfo {author} {\bibfnamefont {M.}~\bibnamefont {Turner}}, \
  and\ \bibinfo {author} {\bibfnamefont {D.}~\bibnamefont {Huterer}},\ }\href
  {\doibase 10.1146/annurev.astro.46.060407.145243} {\bibfield  {journal}
  {\bibinfo  {journal} {Ann.Rev.Astron.Astrophys.}\ }\textbf {\bibinfo {volume}
  {46}},\ \bibinfo {pages} {385} (\bibinfo {year} {2008})},\ \Eprint
  {http://arxiv.org/abs/0803.0982} {arXiv:0803.0982 [astro-ph]} \BibitemShut
  {NoStop}%
\bibitem [{\citenamefont {Weinberg}\ \emph {et~al.}(2013)\citenamefont
  {Weinberg}, \citenamefont {Mortonson}, \citenamefont {Eisenstein},
  \citenamefont {Hirata}, \citenamefont {Riess} \emph
  {et~al.}}]{Weinberg:2012es}%
  \BibitemOpen
  \bibfield  {author} {\bibinfo {author} {\bibfnamefont {D.~H.}\ \bibnamefont
  {Weinberg}}, \bibinfo {author} {\bibfnamefont {M.~J.}\ \bibnamefont
  {Mortonson}}, \bibinfo {author} {\bibfnamefont {D.~J.}\ \bibnamefont
  {Eisenstein}}, \bibinfo {author} {\bibfnamefont {C.}~\bibnamefont {Hirata}},
  \bibinfo {author} {\bibfnamefont {A.~G.}\ \bibnamefont {Riess}},  \emph
  {et~al.},\ }\href {\doibase 10.1016/j.physrep.2013.05.001} {\bibfield
  {journal} {\bibinfo  {journal} {Phys.Rept.}\ }\textbf {\bibinfo {volume}
  {530}},\ \bibinfo {pages} {87} (\bibinfo {year} {2013})},\ \Eprint
  {http://arxiv.org/abs/1201.2434} {arXiv:1201.2434 [astro-ph.CO]} \BibitemShut
  {NoStop}%
\bibitem [{\citenamefont {Riess}\ \emph {et~al.}(1998)\citenamefont {Riess}
  \emph {et~al.}}]{Riess:1998cb}%
  \BibitemOpen
  \bibfield  {author} {\bibinfo {author} {\bibfnamefont {A.~G.}\ \bibnamefont
  {Riess}} \emph {et~al.} (\bibinfo {collaboration} {Supernova Search Team}),\
  }\href {\doibase 10.1086/300499} {\bibfield  {journal} {\bibinfo  {journal}
  {Astron.J.}\ }\textbf {\bibinfo {volume} {116}},\ \bibinfo {pages} {1009}
  (\bibinfo {year} {1998})},\ \Eprint {http://arxiv.org/abs/astro-ph/9805201}
  {arXiv:astro-ph/9805201 [astro-ph]} \BibitemShut {NoStop}%
\bibitem [{\citenamefont {Perlmutter}\ \emph {et~al.}(1999)\citenamefont
  {Perlmutter} \emph {et~al.}}]{Perlmutter:1998np}%
  \BibitemOpen
  \bibfield  {author} {\bibinfo {author} {\bibfnamefont {S.}~\bibnamefont
  {Perlmutter}} \emph {et~al.} (\bibinfo {collaboration} {Supernova Cosmology
  Project}),\ }\href {\doibase 10.1086/307221} {\bibfield  {journal} {\bibinfo
  {journal} {Astrophys.J.}\ }\textbf {\bibinfo {volume} {517}},\ \bibinfo
  {pages} {565} (\bibinfo {year} {1999})},\ \Eprint
  {http://arxiv.org/abs/astro-ph/9812133} {arXiv:astro-ph/9812133 [astro-ph]}
  \BibitemShut {NoStop}%
\bibitem [{\citenamefont {Guy}\ \emph {et~al.}(2007)\citenamefont {Guy},
  \citenamefont {Astier}, \citenamefont {Baumont}, \citenamefont {Hardin},
  \citenamefont {Pain} \emph {et~al.}}]{Guy:2007dv}%
  \BibitemOpen
  \bibfield  {author} {\bibinfo {author} {\bibfnamefont {J.}~\bibnamefont
  {Guy}}, \bibinfo {author} {\bibfnamefont {P.}~\bibnamefont {Astier}},
  \bibinfo {author} {\bibfnamefont {S.}~\bibnamefont {Baumont}}, \bibinfo
  {author} {\bibfnamefont {D.}~\bibnamefont {Hardin}}, \bibinfo {author}
  {\bibfnamefont {R.}~\bibnamefont {Pain}},  \emph {et~al.},\ }\href {\doibase
  10.1051/0004-6361:20066930} {\bibfield  {journal} {\bibinfo  {journal}
  {Astron.Astrophys.}\ }\textbf {\bibinfo {volume} {466}},\ \bibinfo {pages}
  {11} (\bibinfo {year} {2007})},\ \Eprint
  {http://arxiv.org/abs/astro-ph/0701828} {arXiv:astro-ph/0701828 [ASTRO-PH]}
  \BibitemShut {NoStop}%
\bibitem [{\citenamefont {Conley}\ \emph {et~al.}(2008)\citenamefont {Conley},
  \citenamefont {Sullivan}, \citenamefont {Hsiao}, \citenamefont {Guy},
  \citenamefont {Astier} \emph {et~al.}}]{Conley:2008xx}%
  \BibitemOpen
  \bibfield  {author} {\bibinfo {author} {\bibfnamefont {A.~J.}\ \bibnamefont
  {Conley}}, \bibinfo {author} {\bibfnamefont {M.}~\bibnamefont {Sullivan}},
  \bibinfo {author} {\bibfnamefont {E.}~\bibnamefont {Hsiao}}, \bibinfo
  {author} {\bibfnamefont {J.}~\bibnamefont {Guy}}, \bibinfo {author}
  {\bibfnamefont {P.}~\bibnamefont {Astier}},  \emph {et~al.},\ }\href@noop {}
  {\  (\bibinfo {year} {2008})},\ \Eprint {http://arxiv.org/abs/0803.3441}
  {arXiv:0803.3441 [astro-ph]} \BibitemShut {NoStop}%
\bibitem [{\citenamefont {Ruiz}\ \emph {et~al.}(2012)\citenamefont {Ruiz},
  \citenamefont {Shafer}, \citenamefont {Huterer},\ and\ \citenamefont
  {Conley}}]{Ruiz:2012rc}%
  \BibitemOpen
  \bibfield  {author} {\bibinfo {author} {\bibfnamefont {E.~J.}\ \bibnamefont
  {Ruiz}}, \bibinfo {author} {\bibfnamefont {D.~L.}\ \bibnamefont {Shafer}},
  \bibinfo {author} {\bibfnamefont {D.}~\bibnamefont {Huterer}}, \ and\
  \bibinfo {author} {\bibfnamefont {A.}~\bibnamefont {Conley}},\ }\href
  {\doibase 10.1103/PhysRevD.86.103004} {\bibfield  {journal} {\bibinfo
  {journal} {Phys.Rev.}\ }\textbf {\bibinfo {volume} {D86}},\ \bibinfo {pages}
  {103004} (\bibinfo {year} {2012})},\ \Eprint {http://arxiv.org/abs/1207.4781}
  {arXiv:1207.4781 [astro-ph.CO]} \BibitemShut {NoStop}%
\bibitem [{\citenamefont {Conley}\ \emph {et~al.}(2011)\citenamefont {Conley},
  \citenamefont {Guy}, \citenamefont {Sullivan}, \citenamefont {Regnault},
  \citenamefont {Astier} \emph {et~al.}}]{Conley:2011ku}%
  \BibitemOpen
  \bibfield  {author} {\bibinfo {author} {\bibfnamefont {A.}~\bibnamefont
  {Conley}}, \bibinfo {author} {\bibfnamefont {J.}~\bibnamefont {Guy}},
  \bibinfo {author} {\bibfnamefont {M.}~\bibnamefont {Sullivan}}, \bibinfo
  {author} {\bibfnamefont {N.}~\bibnamefont {Regnault}}, \bibinfo {author}
  {\bibfnamefont {P.}~\bibnamefont {Astier}},  \emph {et~al.},\ }\href
  {\doibase 10.1088/0067-0049/192/1/1} {\bibfield  {journal} {\bibinfo
  {journal} {Astrophys.J.Suppl.}\ }\textbf {\bibinfo {volume} {192}},\ \bibinfo
  {pages} {1} (\bibinfo {year} {2011})},\ \Eprint
  {http://arxiv.org/abs/1104.1443} {arXiv:1104.1443 [astro-ph.CO]} \BibitemShut
  {NoStop}%
\bibitem [{\citenamefont {Betoule}\ \emph {et~al.}(2013)\citenamefont {Betoule}
  \emph {et~al.}}]{Betoule:2012an}%
  \BibitemOpen
  \bibfield  {author} {\bibinfo {author} {\bibfnamefont {M.}~\bibnamefont
  {Betoule}} \emph {et~al.} (\bibinfo {collaboration} {SDSS Collaboration}),\
  }\href {\doibase 10.1051/0004-6361/201220610} {\bibfield  {journal} {\bibinfo
   {journal} {Astron.Astrophys.}\ }\textbf {\bibinfo {volume} {552}},\ \bibinfo
  {pages} {124} (\bibinfo {year} {2013})},\ \Eprint
  {http://arxiv.org/abs/1212.4864} {arXiv:1212.4864 [astro-ph.CO]} \BibitemShut
  {NoStop}%
\bibitem [{\citenamefont {Beutler}\ \emph {et~al.}(2011)\citenamefont
  {Beutler}, \citenamefont {Blake}, \citenamefont {Colless}, \citenamefont
  {Jones}, \citenamefont {Staveley-Smith} \emph {et~al.}}]{Beutler:2011hx}%
  \BibitemOpen
  \bibfield  {author} {\bibinfo {author} {\bibfnamefont {F.}~\bibnamefont
  {Beutler}}, \bibinfo {author} {\bibfnamefont {C.}~\bibnamefont {Blake}},
  \bibinfo {author} {\bibfnamefont {M.}~\bibnamefont {Colless}}, \bibinfo
  {author} {\bibfnamefont {D.~H.}\ \bibnamefont {Jones}}, \bibinfo {author}
  {\bibfnamefont {L.}~\bibnamefont {Staveley-Smith}},  \emph {et~al.},\ }\href
  {\doibase 10.1111/j.1365-2966.2011.19250.x} {\bibfield  {journal} {\bibinfo
  {journal} {Mon.Not.Roy.Astron.Soc.}\ }\textbf {\bibinfo {volume} {416}},\
  \bibinfo {pages} {3017} (\bibinfo {year} {2011})},\ \Eprint
  {http://arxiv.org/abs/1106.3366} {arXiv:1106.3366 [astro-ph.CO]} \BibitemShut
  {NoStop}%
\bibitem [{\citenamefont {Padmanabhan}\ \emph {et~al.}(2012)\citenamefont
  {Padmanabhan}, \citenamefont {Xu}, \citenamefont {Eisenstein}, \citenamefont
  {Scalzo}, \citenamefont {Cuesta} \emph {et~al.}}]{Padmanabhan:2012hf}%
  \BibitemOpen
  \bibfield  {author} {\bibinfo {author} {\bibfnamefont {N.}~\bibnamefont
  {Padmanabhan}}, \bibinfo {author} {\bibfnamefont {X.}~\bibnamefont {Xu}},
  \bibinfo {author} {\bibfnamefont {D.~J.}\ \bibnamefont {Eisenstein}},
  \bibinfo {author} {\bibfnamefont {R.}~\bibnamefont {Scalzo}}, \bibinfo
  {author} {\bibfnamefont {A.~J.}\ \bibnamefont {Cuesta}},  \emph {et~al.},\
  }\href {\doibase 10.1111/j.1365-2966.2012.21888.x} {\bibfield  {journal}
  {\bibinfo  {journal} {Mon.Not.Roy.Astron.Soc.}\ }\textbf {\bibinfo {volume}
  {427}},\ \bibinfo {pages} {2132} (\bibinfo {year} {2012})},\ \Eprint
  {http://arxiv.org/abs/1202.0090} {arXiv:1202.0090 [astro-ph.CO]} \BibitemShut
  {NoStop}%
\bibitem [{\citenamefont {Anderson}\ \emph {et~al.}(2013)\citenamefont
  {Anderson}, \citenamefont {Aubourg}, \citenamefont {Bailey}, \citenamefont
  {Bizyaev}, \citenamefont {Blanton} \emph {et~al.}}]{Anderson:2012sa}%
  \BibitemOpen
  \bibfield  {author} {\bibinfo {author} {\bibfnamefont {L.}~\bibnamefont
  {Anderson}}, \bibinfo {author} {\bibfnamefont {E.}~\bibnamefont {Aubourg}},
  \bibinfo {author} {\bibfnamefont {S.}~\bibnamefont {Bailey}}, \bibinfo
  {author} {\bibfnamefont {D.}~\bibnamefont {Bizyaev}}, \bibinfo {author}
  {\bibfnamefont {M.}~\bibnamefont {Blanton}},  \emph {et~al.},\ }\href
  {\doibase 10.1111/j.1365-2966.2012.22066.x} {\bibfield  {journal} {\bibinfo
  {journal} {Mon.Not.Roy.Astron.Soc.}\ }\textbf {\bibinfo {volume} {427}},\
  \bibinfo {pages} {3435} (\bibinfo {year} {2013})},\ \Eprint
  {http://arxiv.org/abs/1203.6594} {arXiv:1203.6594 [astro-ph.CO]} \BibitemShut
  {NoStop}%
\bibitem [{\citenamefont {Eisenstein}\ \emph {et~al.}(2005)\citenamefont
  {Eisenstein} \emph {et~al.}}]{Eisenstein:2005su}%
  \BibitemOpen
  \bibfield  {author} {\bibinfo {author} {\bibfnamefont {D.~J.}\ \bibnamefont
  {Eisenstein}} \emph {et~al.} (\bibinfo {collaboration} {SDSS
  Collaboration}),\ }\href {\doibase 10.1086/466512} {\bibfield  {journal}
  {\bibinfo  {journal} {Astrophys.J.}\ }\textbf {\bibinfo {volume} {633}},\
  \bibinfo {pages} {560} (\bibinfo {year} {2005})},\ \Eprint
  {http://arxiv.org/abs/astro-ph/0501171} {arXiv:astro-ph/0501171 [astro-ph]}
  \BibitemShut {NoStop}%
\bibitem [{\citenamefont {Eisenstein}\ and\ \citenamefont
  {Hu}(1998)}]{Eisenstein:1997ik}%
  \BibitemOpen
  \bibfield  {author} {\bibinfo {author} {\bibfnamefont {D.~J.}\ \bibnamefont
  {Eisenstein}}\ and\ \bibinfo {author} {\bibfnamefont {W.}~\bibnamefont
  {Hu}},\ }\href {\doibase 10.1086/305424} {\bibfield  {journal} {\bibinfo
  {journal} {Astrophys.J.}\ }\textbf {\bibinfo {volume} {496}},\ \bibinfo
  {pages} {605} (\bibinfo {year} {1998})},\ \Eprint
  {http://arxiv.org/abs/astro-ph/9709112} {arXiv:astro-ph/9709112 [astro-ph]}
  \BibitemShut {NoStop}%
\bibitem [{\citenamefont {Percival}\ \emph {et~al.}(2010)\citenamefont
  {Percival} \emph {et~al.}}]{Percival:2009xn}%
  \BibitemOpen
  \bibfield  {author} {\bibinfo {author} {\bibfnamefont {W.~J.}\ \bibnamefont
  {Percival}} \emph {et~al.} (\bibinfo {collaboration} {SDSS Collaboration}),\
  }\href {\doibase 10.1111/j.1365-2966.2009.15812.x} {\bibfield  {journal}
  {\bibinfo  {journal} {Mon.Not.Roy.Astron.Soc.}\ }\textbf {\bibinfo {volume}
  {401}},\ \bibinfo {pages} {2148} (\bibinfo {year} {2010})},\ \Eprint
  {http://arxiv.org/abs/0907.1660} {arXiv:0907.1660 [astro-ph.CO]} \BibitemShut
  {NoStop}%
\bibitem [{\citenamefont {Blake}\ \emph {et~al.}(2011)\citenamefont {Blake},
  \citenamefont {Kazin}, \citenamefont {Beutler}, \citenamefont {Davis},
  \citenamefont {Parkinson} \emph {et~al.}}]{Blake:2011en}%
  \BibitemOpen
  \bibfield  {author} {\bibinfo {author} {\bibfnamefont {C.}~\bibnamefont
  {Blake}}, \bibinfo {author} {\bibfnamefont {E.}~\bibnamefont {Kazin}},
  \bibinfo {author} {\bibfnamefont {F.}~\bibnamefont {Beutler}}, \bibinfo
  {author} {\bibfnamefont {T.}~\bibnamefont {Davis}}, \bibinfo {author}
  {\bibfnamefont {D.}~\bibnamefont {Parkinson}},  \emph {et~al.},\ }\href
  {\doibase 10.1111/j.1365-2966.2011.19592.x} {\bibfield  {journal} {\bibinfo
  {journal} {Mon.Not.Roy.Astron.Soc.}\ }\textbf {\bibinfo {volume} {418}},\
  \bibinfo {pages} {1707} (\bibinfo {year} {2011})},\ \Eprint
  {http://arxiv.org/abs/1108.2635} {arXiv:1108.2635 [astro-ph.CO]} \BibitemShut
  {NoStop}%
\bibitem [{\citenamefont {Frieman}\ \emph {et~al.}(2003)\citenamefont
  {Frieman}, \citenamefont {Huterer}, \citenamefont {Linder},\ and\
  \citenamefont {Turner}}]{Frieman:2002wi}%
  \BibitemOpen
  \bibfield  {author} {\bibinfo {author} {\bibfnamefont {J.~A.}\ \bibnamefont
  {Frieman}}, \bibinfo {author} {\bibfnamefont {D.}~\bibnamefont {Huterer}},
  \bibinfo {author} {\bibfnamefont {E.~V.}\ \bibnamefont {Linder}}, \ and\
  \bibinfo {author} {\bibfnamefont {M.~S.}\ \bibnamefont {Turner}},\ }\href
  {\doibase 10.1103/PhysRevD.67.083505} {\bibfield  {journal} {\bibinfo
  {journal} {Phys.Rev.}\ }\textbf {\bibinfo {volume} {D67}},\ \bibinfo {pages}
  {083505} (\bibinfo {year} {2003})},\ \Eprint
  {http://arxiv.org/abs/astro-ph/0208100} {arXiv:astro-ph/0208100 [astro-ph]}
  \BibitemShut {NoStop}%
\bibitem [{\citenamefont {Hinshaw}\ \emph {et~al.}(2013)\citenamefont {Hinshaw}
  \emph {et~al.}}]{Hinshaw:2012aka}%
  \BibitemOpen
  \bibfield  {author} {\bibinfo {author} {\bibfnamefont {G.}~\bibnamefont
  {Hinshaw}} \emph {et~al.} (\bibinfo {collaboration} {WMAP}),\ }\href
  {\doibase 10.1088/0067-0049/208/2/19} {\bibfield  {journal} {\bibinfo
  {journal} {Astrophys.J.Suppl.}\ }\textbf {\bibinfo {volume} {208}},\ \bibinfo
  {pages} {19} (\bibinfo {year} {2013})},\ \Eprint
  {http://arxiv.org/abs/1212.5226} {arXiv:1212.5226 [astro-ph.CO]} \BibitemShut
  {NoStop}%
\bibitem [{\citenamefont {Hu}\ and\ \citenamefont
  {Sugiyama}(1996)}]{Hu:1995en}%
  \BibitemOpen
  \bibfield  {author} {\bibinfo {author} {\bibfnamefont {W.}~\bibnamefont
  {Hu}}\ and\ \bibinfo {author} {\bibfnamefont {N.}~\bibnamefont {Sugiyama}},\
  }\href {\doibase 10.1086/177989} {\bibfield  {journal} {\bibinfo  {journal}
  {Astrophys.J.}\ }\textbf {\bibinfo {volume} {471}},\ \bibinfo {pages} {542}
  (\bibinfo {year} {1996})},\ \Eprint {http://arxiv.org/abs/astro-ph/9510117}
  {arXiv:astro-ph/9510117 [astro-ph]} \BibitemShut {NoStop}%
\bibitem [{\citenamefont {Wang}\ and\ \citenamefont
  {Wang}(2013{\natexlab{a}})}]{Wang:2013mha}%
  \BibitemOpen
  \bibfield  {author} {\bibinfo {author} {\bibfnamefont {Y.}~\bibnamefont
  {Wang}}\ and\ \bibinfo {author} {\bibfnamefont {S.}~\bibnamefont {Wang}},\
  }\href {\doibase 10.1103/PhysRevD.88.043522} {\bibfield  {journal} {\bibinfo
  {journal} {Phys.Rev.}\ }\textbf {\bibinfo {volume} {D88}},\ \bibinfo {pages}
  {043522} (\bibinfo {year} {2013}{\natexlab{a}})},\ \Eprint
  {http://arxiv.org/abs/1304.4514} {arXiv:1304.4514 [astro-ph.CO]} \BibitemShut
  {NoStop}%
\bibitem [{\citenamefont {Rigault}\ \emph {et~al.}(2013)\citenamefont
  {Rigault}, \citenamefont {Copin}, \citenamefont {Aldering}, \citenamefont
  {Antilogus}, \citenamefont {Aragon} \emph {et~al.}}]{Rigault:2013gux}%
  \BibitemOpen
  \bibfield  {author} {\bibinfo {author} {\bibfnamefont {M.}~\bibnamefont
  {Rigault}}, \bibinfo {author} {\bibfnamefont {Y.}~\bibnamefont {Copin}},
  \bibinfo {author} {\bibfnamefont {G.}~\bibnamefont {Aldering}}, \bibinfo
  {author} {\bibfnamefont {P.}~\bibnamefont {Antilogus}}, \bibinfo {author}
  {\bibfnamefont {C.}~\bibnamefont {Aragon}},  \emph {et~al.},\ }\href
  {\doibase 10.1051/0004-6361/201322104} {\  (\bibinfo {year} {2013}),\
  10.1051/0004-6361/201322104},\ \Eprint {http://arxiv.org/abs/1309.1182}
  {arXiv:1309.1182 [astro-ph.CO]} \BibitemShut {NoStop}%
\bibitem [{\citenamefont {Kelly}(2007)}]{Kelly:2007jy}%
  \BibitemOpen
  \bibfield  {author} {\bibinfo {author} {\bibfnamefont {B.~C.}\ \bibnamefont
  {Kelly}},\ }\href {\doibase 10.1086/519947} {\bibfield  {journal} {\bibinfo
  {journal} {Astrophys.J.}\ }\textbf {\bibinfo {volume} {665}},\ \bibinfo
  {pages} {1489} (\bibinfo {year} {2007})},\ \Eprint
  {http://arxiv.org/abs/0705.2774} {arXiv:0705.2774 [astro-ph]} \BibitemShut
  {NoStop}%
\bibitem [{\citenamefont {Sullivan}\ \emph {et~al.}(2010)\citenamefont
  {Sullivan}, \citenamefont {Conley}, \citenamefont {Howell}, \citenamefont
  {Neill}, \citenamefont {Astier} \emph {et~al.}}]{Sullivan:2010mg}%
  \BibitemOpen
  \bibfield  {author} {\bibinfo {author} {\bibfnamefont {M.}~\bibnamefont
  {Sullivan}}, \bibinfo {author} {\bibfnamefont {A.}~\bibnamefont {Conley}},
  \bibinfo {author} {\bibfnamefont {D.}~\bibnamefont {Howell}}, \bibinfo
  {author} {\bibfnamefont {J.}~\bibnamefont {Neill}}, \bibinfo {author}
  {\bibfnamefont {P.}~\bibnamefont {Astier}},  \emph {et~al.},\ }\href@noop {}
  {\bibfield  {journal} {\bibinfo  {journal} {Mon.Not.Roy.Astron.Soc.}\
  }\textbf {\bibinfo {volume} {406}},\ \bibinfo {pages} {782} (\bibinfo {year}
  {2010})},\ \Eprint {http://arxiv.org/abs/1003.5119} {arXiv:1003.5119
  [astro-ph.CO]} \BibitemShut {NoStop}%
\bibitem [{\citenamefont {Gupta}\ \emph {et~al.}(2011)\citenamefont {Gupta},
  \citenamefont {D'Andrea}, \citenamefont {Sako}, \citenamefont {Conroy},
  \citenamefont {Smith} \emph {et~al.}}]{Gupta:2011pa}%
  \BibitemOpen
  \bibfield  {author} {\bibinfo {author} {\bibfnamefont {R.~R.}\ \bibnamefont
  {Gupta}}, \bibinfo {author} {\bibfnamefont {C.~B.}\ \bibnamefont {D'Andrea}},
  \bibinfo {author} {\bibfnamefont {M.}~\bibnamefont {Sako}}, \bibinfo {author}
  {\bibfnamefont {C.}~\bibnamefont {Conroy}}, \bibinfo {author} {\bibfnamefont
  {M.}~\bibnamefont {Smith}},  \emph {et~al.},\ }\href {\doibase
  10.1088/0004-637X/741/2/127, 10.1088/0004-637X/740/2/92} {\bibfield
  {journal} {\bibinfo  {journal} {Astrophys.J.}\ }\textbf {\bibinfo {volume}
  {740}},\ \bibinfo {pages} {92} (\bibinfo {year} {2011})},\ \Eprint
  {http://arxiv.org/abs/1107.6003} {arXiv:1107.6003 [astro-ph.CO]} \BibitemShut
  {NoStop}%
\bibitem [{\citenamefont {Kessler}\ \emph {et~al.}(2013)\citenamefont
  {Kessler}, \citenamefont {Guy}, \citenamefont {Marriner}, \citenamefont
  {Betoule}, \citenamefont {Brinkmann} \emph {et~al.}}]{Kessler:2012gn}%
  \BibitemOpen
  \bibfield  {author} {\bibinfo {author} {\bibfnamefont {R.}~\bibnamefont
  {Kessler}}, \bibinfo {author} {\bibfnamefont {J.}~\bibnamefont {Guy}},
  \bibinfo {author} {\bibfnamefont {J.}~\bibnamefont {Marriner}}, \bibinfo
  {author} {\bibfnamefont {M.}~\bibnamefont {Betoule}}, \bibinfo {author}
  {\bibfnamefont {J.}~\bibnamefont {Brinkmann}},  \emph {et~al.},\ }\href
  {\doibase 10.1088/0004-637X/764/1/48} {\bibfield  {journal} {\bibinfo
  {journal} {Astrophys.J.}\ }\textbf {\bibinfo {volume} {764}},\ \bibinfo
  {pages} {48} (\bibinfo {year} {2013})},\ \Eprint
  {http://arxiv.org/abs/1209.2482} {arXiv:1209.2482 [astro-ph.CO]} \BibitemShut
  {NoStop}%
\bibitem [{\citenamefont {Johansson}\ \emph {et~al.}(2012)\citenamefont
  {Johansson}, \citenamefont {Thomas}, \citenamefont {Pforr}, \citenamefont
  {Maraston}, \citenamefont {Nichol} \emph {et~al.}}]{Johansson:2012si}%
  \BibitemOpen
  \bibfield  {author} {\bibinfo {author} {\bibfnamefont {J.}~\bibnamefont
  {Johansson}}, \bibinfo {author} {\bibfnamefont {D.}~\bibnamefont {Thomas}},
  \bibinfo {author} {\bibfnamefont {J.}~\bibnamefont {Pforr}}, \bibinfo
  {author} {\bibfnamefont {C.}~\bibnamefont {Maraston}}, \bibinfo {author}
  {\bibfnamefont {R.~C.}\ \bibnamefont {Nichol}},  \emph {et~al.},\ }\href@noop
  {} {\  (\bibinfo {year} {2012})},\ \Eprint {http://arxiv.org/abs/1211.1386}
  {arXiv:1211.1386 [astro-ph.CO]} \BibitemShut {NoStop}%
\bibitem [{\citenamefont {Childress}\ \emph {et~al.}(2013)\citenamefont
  {Childress}, \citenamefont {Aldering}, \citenamefont {Antilogus},
  \citenamefont {Aragon}, \citenamefont {Bailey} \emph
  {et~al.}}]{Childress:2013xna}%
  \BibitemOpen
  \bibfield  {author} {\bibinfo {author} {\bibfnamefont {M.}~\bibnamefont
  {Childress}}, \bibinfo {author} {\bibfnamefont {G.}~\bibnamefont {Aldering}},
  \bibinfo {author} {\bibfnamefont {P.}~\bibnamefont {Antilogus}}, \bibinfo
  {author} {\bibfnamefont {C.}~\bibnamefont {Aragon}}, \bibinfo {author}
  {\bibfnamefont {S.}~\bibnamefont {Bailey}},  \emph {et~al.},\ }\href
  {\doibase 10.1088/0004-637X/770/2/108} {\bibfield  {journal} {\bibinfo
  {journal} {Astrophys.J.}\ }\textbf {\bibinfo {volume} {770}},\ \bibinfo
  {pages} {108} (\bibinfo {year} {2013})},\ \Eprint
  {http://arxiv.org/abs/1304.4720} {arXiv:1304.4720 [astro-ph.CO]} \BibitemShut
  {NoStop}%
\bibitem [{\citenamefont {Wang}\ and\ \citenamefont
  {Wang}(2013{\natexlab{b}})}]{Wang:2013yja}%
  \BibitemOpen
  \bibfield  {author} {\bibinfo {author} {\bibfnamefont {S.}~\bibnamefont
  {Wang}}\ and\ \bibinfo {author} {\bibfnamefont {Y.}~\bibnamefont {Wang}},\
  }\href {\doibase 10.1103/PhysRevD.88.043511} {\bibfield  {journal} {\bibinfo
  {journal} {Phys.Rev.}\ }\textbf {\bibinfo {volume} {D88}},\ \bibinfo {pages}
  {043511} (\bibinfo {year} {2013}{\natexlab{b}})},\ \Eprint
  {http://arxiv.org/abs/1306.6423} {arXiv:1306.6423 [astro-ph.CO]} \BibitemShut
  {NoStop}%
\bibitem [{\citenamefont {Riess}\ \emph {et~al.}(2011)\citenamefont {Riess},
  \citenamefont {Macri}, \citenamefont {Casertano}, \citenamefont {Lampeitl},
  \citenamefont {Ferguson} \emph {et~al.}}]{Riess:2011yx}%
  \BibitemOpen
  \bibfield  {author} {\bibinfo {author} {\bibfnamefont {A.~G.}\ \bibnamefont
  {Riess}}, \bibinfo {author} {\bibfnamefont {L.}~\bibnamefont {Macri}},
  \bibinfo {author} {\bibfnamefont {S.}~\bibnamefont {Casertano}}, \bibinfo
  {author} {\bibfnamefont {H.}~\bibnamefont {Lampeitl}}, \bibinfo {author}
  {\bibfnamefont {H.~C.}\ \bibnamefont {Ferguson}},  \emph {et~al.},\ }\href
  {\doibase 10.1088/0004-637X/732/2/129, 10.1088/0004-637X/730/2/119}
  {\bibfield  {journal} {\bibinfo  {journal} {Astrophys.J.}\ }\textbf {\bibinfo
  {volume} {730}},\ \bibinfo {pages} {119} (\bibinfo {year} {2011})},\ \Eprint
  {http://arxiv.org/abs/1103.2976} {arXiv:1103.2976 [astro-ph.CO]} \BibitemShut
  {NoStop}%
\bibitem [{\citenamefont {Hu}(2005)}]{Hu:2004kn}%
  \BibitemOpen
  \bibfield  {author} {\bibinfo {author} {\bibfnamefont {W.}~\bibnamefont
  {Hu}},\ }\href@noop {} {\bibfield  {journal} {\bibinfo  {journal} {ASP
  Conf.Ser.}\ }\textbf {\bibinfo {volume} {339}},\ \bibinfo {pages} {215}
  (\bibinfo {year} {2005})},\ \Eprint {http://arxiv.org/abs/astro-ph/0407158}
  {arXiv:astro-ph/0407158 [astro-ph]} \BibitemShut {NoStop}%
\bibitem [{\citenamefont {Efstathiou}(2013)}]{Efstathiou:2013via}%
  \BibitemOpen
  \bibfield  {author} {\bibinfo {author} {\bibfnamefont {G.}~\bibnamefont
  {Efstathiou}},\ }\href@noop {} {\  (\bibinfo {year} {2013})},\ \Eprint
  {http://arxiv.org/abs/1311.3461} {arXiv:1311.3461 [astro-ph.CO]} \BibitemShut
  {NoStop}%
\bibitem [{\citenamefont {Marra}\ \emph {et~al.}(2013)\citenamefont {Marra},
  \citenamefont {Amendola}, \citenamefont {Sawicki},\ and\ \citenamefont
  {Valkenburg}}]{Marra:2013rba}%
  \BibitemOpen
  \bibfield  {author} {\bibinfo {author} {\bibfnamefont {V.}~\bibnamefont
  {Marra}}, \bibinfo {author} {\bibfnamefont {L.}~\bibnamefont {Amendola}},
  \bibinfo {author} {\bibfnamefont {I.}~\bibnamefont {Sawicki}}, \ and\
  \bibinfo {author} {\bibfnamefont {W.}~\bibnamefont {Valkenburg}},\ }\href
  {\doibase 10.1103/PhysRevLett.110.241305} {\bibfield  {journal} {\bibinfo
  {journal} {Phys.Rev.Lett.}\ }\textbf {\bibinfo {volume} {110}},\ \bibinfo
  {pages} {241305} (\bibinfo {year} {2013})},\ \Eprint
  {http://arxiv.org/abs/1303.3121} {arXiv:1303.3121 [astro-ph.CO]} \BibitemShut
  {NoStop}%
\bibitem [{\citenamefont {Macaulay}\ \emph {et~al.}(2013)\citenamefont
  {Macaulay}, \citenamefont {Wehus},\ and\ \citenamefont
  {Eriksen}}]{Macaulay:2013swa}%
  \BibitemOpen
  \bibfield  {author} {\bibinfo {author} {\bibfnamefont {E.}~\bibnamefont
  {Macaulay}}, \bibinfo {author} {\bibfnamefont {I.~K.}\ \bibnamefont {Wehus}},
  \ and\ \bibinfo {author} {\bibfnamefont {H.~K.}\ \bibnamefont {Eriksen}},\
  }\href@noop {} {\  (\bibinfo {year} {2013})},\ \Eprint
  {http://arxiv.org/abs/1303.6583} {arXiv:1303.6583 [astro-ph.CO]} \BibitemShut
  {NoStop}%
\bibitem [{\citenamefont {Spergel}\ \emph {et~al.}(2013)\citenamefont
  {Spergel}, \citenamefont {Flauger},\ and\ \citenamefont
  {Hlozek}}]{Spergel:2013rxa}%
  \BibitemOpen
  \bibfield  {author} {\bibinfo {author} {\bibfnamefont {D.}~\bibnamefont
  {Spergel}}, \bibinfo {author} {\bibfnamefont {R.}~\bibnamefont {Flauger}}, \
  and\ \bibinfo {author} {\bibfnamefont {R.}~\bibnamefont {Hlozek}},\
  }\href@noop {} {\  (\bibinfo {year} {2013})},\ \Eprint
  {http://arxiv.org/abs/1312.3313} {arXiv:1312.3313 [astro-ph.CO]} \BibitemShut
  {NoStop}%
\end{thebibliography}%

\end{document}